\documentclass[letterpaper]{iopart}
\usepackage{iopams,setstack,amsopn} 
\expandafter\let\csname equation*\endcsname\relax
\expandafter\let\csname endequation*\endcsname\relax
\usepackage{amsmath}
\usepackage{bm}
\usepackage{physics}
\usepackage{cite}
\usepackage{dsfont}

\usepackage[varg]{txfonts}
\usepackage{graphicx}
\usepackage{color}
\usepackage[colorlinks=true,citecolor=blue]{hyperref}
\usepackage[margin=1in]{geometry}
\usepackage[caption=false]{subfig}
\usepackage{booktabs}
\usepackage{cleveref}
\usepackage{comment}

\crefname{equation}{Eq.}{Eqs.}
\Crefname{equation}{Equation}{Equations}
\crefname{figure}{Fig.}{Figs.}
\Crefname{figure}{Figure}{Figures}
\crefname{section}{Sect.}{Sects.}
\Crefname{section}{Section}{Sections}
\crefname{table}{Table}{Tables}
\crefname{appendix}{}{}

\usepackage[stable]{footmisc}

\usepackage[normalem]{ulem}
\usepackage{soul}

\usepackage{multirow}

\newcommand{\be}{\begin{equation}}
\newcommand{\ee}{\end{equation}}

\newcommand{\dg}{^\dagger}

\definecolor{burntorange}{rgb}{0.8, 0.33, 0.0}

\definecolor{adpcolor}{rgb}{0.36,0.54,0.66}

\definecolor{abcolor}{rgb}{0.6, 0.6, 0.1}

\newcommand{\zp}{\mbox{0-$\pi$}}

\begin{document}

\title{Control and Coherence Time Enhancement of the \zp{} Qubit}

\author{Agustin\ Di Paolo$^1$, Arne\ L.\ Grimsmo$^{1,2}$, Peter Groszkowski$^3$\footnote{Present Address: Institute for Molecular Engineering, University of Chicago, Chicago, Illinois 60637, USA.}, Jens Koch$^3$ and Alexandre\ Blais$^{1,4}$}
\address{$^1$ Institut quantique and D\'epartement de Physique, Universit\'e de Sherbrooke, Sherbrooke, QC, Canada.}
\address{$^2$ Centre for Engineered Quantum Systems, School of Physics,The University of Sydney, Sydney, NSW, Australia.}
\address{$^3$ Department of Physics and Astronomy, Northwestern University, Evanston, Illinois 60208, USA.}
\address{$^4$ Canadian Institute for Advanced Re{}search, Toronto, ON, Canada.}
\eads{Agustin.Di.Paolo@USherbrooke.ca}

\begin{abstract}
  Kitaev's \zp{} qubit encodes quantum information in two protected, near-degenerate states of a superconducting quantum circuit. In a recent work, we have shown that the coherence times of a realistic \zp{} device can surpass that of today's best superconducting qubits \cite{groszkowski2017coherence}. Here we address controllability of the \zp{} qubit. Specifically, we investigate the potential for dispersive control and readout, and introduce a new, fast and high-fidelity single-qubit gate that can interpolate smoothly between logical $X$ and $Z$. We characterize the action of this gate using a multi-level treatment of the device, and analyze the impact of circuit element disorder and deviations in control and circuit parameters from their optimal values. Furthermore, we propose a cooling scheme to decrease the photon shot-noise dephasing rate, which we previously found to limit the coherence times of \zp{} devices within reach of current experiments. Using this approach, we predict coherence time enhancements between one and three orders of magnitude, depending on parameter regime.
\end{abstract}

\section{Introduction}
\label{sec:Introduction}

Fault-tolerant quantum computation is likely to require daunting hardware resources \cite{fowler2012surface,reiher2017elucidating}. This fact motivates the search for strategies to reduce the qubit overhead needed for quantum error correction, and drives the development of new quantum error correcting codes \cite{bacon2017sparse,pastawski2015fault,jones2016gauge,tuckett2017ultra,delfosse2017almost}. Furthermore, the reduction of gate errors for physical qubits offers a direct and impactful way of reducing qubit overhead \cite{martinis2015qubit,reiher2017elucidating}. The latter can be achieved both through longer qubit coherence times and better quantum control for gates.

For superconducting circuits, coherence time improvements by as much as five orders of magnitude have been demonstrated \cite{devoret2013superconducting}. This has been possible thanks to advances in several areas, including materials \cite{martinis2005decoherence}, microwave engineering \cite{geerlings2012improving}, shielding \cite{barends2011minimizing,corcoles2011protecting}, and the use of 3D architectures \cite{rigetti2012superconducting,paik2011observation,reagor2013reaching}. Crucially, order-of-magnitude leaps in coherence have also been the result of new qubit designs, such as the transmon and the fluxonium qubits \cite{Koch2007a,manucharyan2009fluxonium}.

In this paper, we consider the superconducting circuit introduced in Ref.~\cite{Brooks2013}, commonly referred to as the \zp{} qubit, and closely related to Kitaev's current mirror proposal \cite{kitaev2006protected}. With a set of non-overlapping logical wavefunctions and very low flux and charge dispersion, the \zp{} qubit displays exponential suppression of relaxation and dephasing. It has been shown that the \zp{} qubit can be used to encode quantum information in a protected subspace \cite{kitaev2006protected,Brooks2013,Dempster2014a,groszkowski2017coherence}, but in a regime of parameters that is challenging to realize with current superconducting quantum circuits. In fact, the fully protected regime of this device exploits a degree of freedom with large quantum fluctuations, something which requires an effective impedance surpassing the quantum of resistance by orders of magnitude. Achieving this regime requires the use of superinductors, which are circuit elements with inductance greater than $\sim100\,\text{nH}$ and with very little stray or ground capacitances \cite{manucharyan2009fluxonium,masluk2012microwave,bell2012quantum,hazard2018nanowire}. We have recently shown that for circuit parameters attainable with \emph{current} superconducting technology, the \zp{} qubit dephasing time is limited by photon shot noise arising from a parasitic circuit mode (which we referred to as the $\zeta$-mode)~\cite{groszkowski2017coherence}. Nevertheless, we found that the \zp{} qubit still has the potential to outperform state-of-the-art superconducting devices. Below, we propose a method to further enhance the coherence time by orders of magnitude by cooling the $\zeta$-mode.

However, as can be expected, the price of intrinsic noise protection in the \zp{} qubit is that it is difficult to perform logical operations on this device. In particular, protection from noise comes in part from the exponentially small overlap of its logical wavefunctions. As a result, matrix elements of local operators between the two logical states will also be small, thus resulting in extremely slow gates. In Ref.~\cite{Brooks2013}, Brooks \textit{et al.} proposed a universal set of protected logical operations based on coupling to an ultra-high impedance LC oscillator. However, these operations were based on an idealized model of the qubit and with parameters that are difficult to realize in practice. Further work is required to determine the potential of this approach in a more realistic setting.

Motivated by the prospect of realizing \zp{} qubits in the near term, we investigate alternative approaches to measurement and control with lower experimental complexity. The operations we propose are not protected in the same sense as those proposed in~\cite{Brooks2013}, because they rely either on operating the device in a regime where the qubit is not fully isolated from the environment, or they make use of excited states outside the qubit manifold. In particular, we develop a single-qubit gate based on a multi-level excursion through higher energy levels. Nevertheless, we hope that these schemes will be useful for both characterization and control of \zp{} qubits in near-to-medium term experiments.

This work is organized as follows. In \cref{sec:nuts}, we introduce the \zp{} qubit and provide a simplified effective model for the \zp{} circuit with only a single degree of freedom. In \cref{sec:coupling}, we discuss general coupling strategies for qubit control and readout, and derive the \zp{} circuit Hamiltonian accounting for stray and parasitic capacitances, disorder in circuit element parameters, as well as coupling to microwave voltage sources and a readout resonator. In \cref{sec:Dispersive}, we analyze dispersive coupling to a resonator, and find that there are regimes of dispersive shift akin to the straddling regime of the transmon qubit~\cite{Koch2007a}. In \cref{sec:QuantumNOTGate}, we introduce a single-qubit gate that achieves population inversion of the \zp{} qubit and can interpolate between logical $X$ and $Z$ smoothly by varying the qubit operation point. Furthermore, we characterize the gate operation as a function of circuit design parameters and analyze its robustness. In \cref{sec:CoolingTheZetaMode}, we propose a method to fight the main qubit dephasing mechanism,  analyze its performance as a function of circuit parameters, and discuss its implementation. We conclude in \cref{sec:Conclusions}.

\section{The \zp{} qubit in a nutshell}
\label{sec:nuts}
In this section we introduce the \zp{} qubit in the ideal case of no circuit element disorder and briefly discuss its properties. In particular, we give an intuitive picture in terms of co-tunneling of Cooper pairs leading to an approximately $\pi$-periodic qubit potential, which is further verified by an effective model accurately describing the low-energy physics of the system.

\subsection{The circuit Hamiltonian}
\label{sssec:Nutshell}

\begin{figure}
    \centering{}
    \includegraphics[scale=.8]{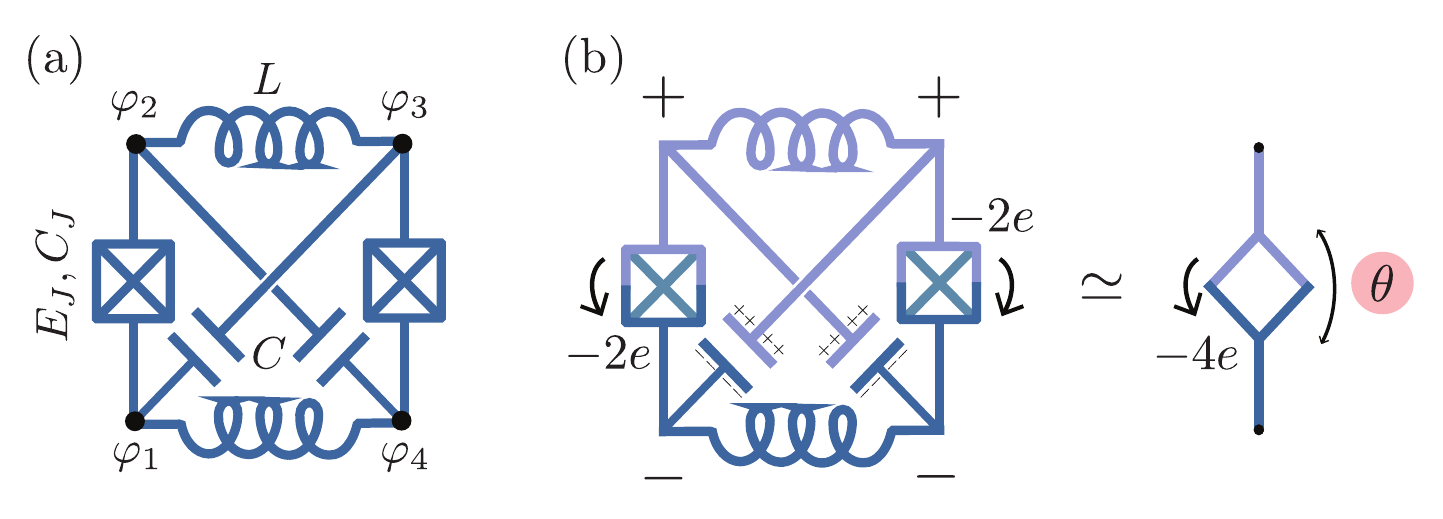}
    \caption{
    The \zp{} qubit in a nutshell. (a) Circuit diagram for the symmetric \zp{} qubit, with pairwise identical circuit elements. (b) Pictorial illustration of co-tunneling of pairs of Cooper pairs across the two junctions, explaining the approximate $\pi$-periodic potential energy, and an equivalent circuit element with only a single degree of freedom $\theta$.}
    \label{fig:nutshell}
\end{figure}

We first consider the symmetric \zp{} circuit, as illustrated in \cref{fig:nutshell} (a), consisting of two Josephson junctions with energy $E_J$, capacitance $C_J$ and plasma frequency $\omega_p=\sqrt{8 E_J E_{C_J}}/\hbar$, two superinductors with inductance $L,$ and two large capacitors with capacitance $C.$ The normal modes of this circuit are 
\begin{equation}
2{\phi} = (\varphi_2-\varphi_3) + (\varphi_4 - \varphi_1),\;\;\; 2{\theta} = (\varphi_2 - \varphi_1) - (\varphi_4 - \varphi_3),\;\;\; 2{\zeta} = (\varphi_2 - \varphi_3) - (\varphi_4 - \varphi_1),\;\;\; 2{\Sigma} = \varphi_1 + \varphi_2 + \varphi_3 + \varphi_4,
\label{eq:normal_modes}
\end{equation}
where $\varphi_i$ is the superconducting phase operator at node $i$ of the circuit. Using these definitions, the symmetric \zp{} qubit Hamiltonian reads \cite{Dempster2014a}
\begin{equation}
H_{0-\pi}^{\text{ideal}}=\frac{q_{\phi}^2}{{2}C_\phi} + \frac{q_{\theta}^2}{{2}C_\theta}-2E_J\cos\theta\cos\qty(\phi-\varphi_{\text{ext}}/2) + E_L \phi^2,
\label{eq:symm_qubit_Hamiltonian}
\end{equation}
where $q_{\phi} = 2e\, n_\phi$ and $q_{\theta} = 2e\,n_\theta$ are the conjugate charge operators associated with $\phi$ and $\theta$ (\emph{i.e.}, $[\phi,n_\phi] = [\theta,n_\theta] = i$) respectively, and $\varphi_{\text{ext}}=\Phi_\text{ext}/\varphi_0$ is the external magnetic flux in units of the reduced flux quantum $\varphi_0 = \hbar/2e.$ Moreover, we have introduced capacitances for the two qubit modes $\phi$ and $\theta$ given by $C_\phi = {2}C_J$ and $C_\theta = {2}(C+C_J)$, respectively, and the inductive energy $E_L = \varphi_0^2/L$.

In the \zp{} qubit, quantum information is stored in the $\{\phi,\theta\}$ degrees of freedom, while $\zeta$ is a spurious low-frequency harmonic mode and $\Sigma$ is a cyclic coordinate. In absence of circuit element disorder, the $\zeta$ and $\Sigma$ modes do not couple to $\phi$ and $\theta$, and are therefore excluded from~\cref{eq:symm_qubit_Hamiltonian}. 

Introducing the effective impedances $Z_{\phi} =\sqrt{{(L/2)}/C_{\phi}}$ and $Z_{\theta} =\sqrt{{(L_J/2)}/C_{\theta}}$, where $L_J = \varphi_0^2/E_J,$ the \zp{} regime is defined by 
\begin{equation}\label{eq:0pi_regime}
Z_{\theta} \ll R_Q \ll Z_{\phi},
\end{equation}
where $R_Q = h/(2e)^2\simeq 6.5\,\text{k}\Omega$ is the superconducting quantum of resistance. We say that a device is in the ``moderate,'' or ``deep'' \zp{} regime, depending on the degree to which the impedance relations are satisfied. The problem of fabricating a qubit in the deep \zp{} regime, includes that of realizing a high-impedance superinductor \cite{manucharyan2012superinductance,Masluk2012,Bell2012}.

\subsection{Exciton tunneling picture}
\label{sssec:QubitStructure}

\Cref{fig:nutshell} (b) shows an approximate equivalence between the \zp{} circuit (to the left) and a circuit element describing tunneling of pairs of Cooper pairs (to the right). The co-tunneling of Cooper pairs or ``exciton'' in the \zp{} circuit can be understood as a consequence of a circuit layout combining branches of superinductors (high impedance) and large capacitances (low impedance). Here, we schematically illustrate how tunneling of a Cooper pair across the left junction of the \zp{} circuit is ``mirrored'' by the simultaneous tunneling of a Cooper pair across the right junction: A Cooper pair tunneling event across the left junction leads to a build up of $-2e$ negative charge on one side of one of the large capacitors, which must be compensated for by a positive charge on the other side. This can happen through a positive $+2e$ Cooper pair tunneling event across the right junction in the opposite  direction.
The co-tunneling of a negative and positive Cooper pair in opposite direction together form an effective exciton tunneling event~\cite{kitaev2006protected}.

Note that no current flows through the superinductors in the limit of $L\to\infty$ ($Z_\phi/R_Q \to \infty$). Superinductors are, however, crucial in defining the non-trivial topology of the circuit, as in their presence we can identify two distinct circuit islands shown as blue (bottom) and violet (top) in~\cref{fig:nutshell} (b). Due to the simultaneous co-tunneling of Cooper pairs across the two junctions, we expect the potential energy to be $\pi$-periodic rather than $2\pi$-periodic in the superconducting phase difference across the two islands, in the limit $L\to \infty$. This expectation can be verified by an effective model for the $\theta$ degree of freedom alone, derived in~\cref{app:1d_effective_Hamiltonian} following a Born-Oppenheimer approach and resulting in the effective Hamiltonian
\begin{equation}
H^{\text{eff}}_{0-\pi} =4 E_{C_{\theta}}(n_{\theta} - n_{g}^\theta)^2 - E_{2}(\varphi_{\text{ext}})\cos2\theta - E_{1}(\varphi_{\text{ext}})\cos\theta,
\label{eq:effective_qubit_Hamiltonian}
\end{equation}
where $E_{C_\theta} = e^2/2C_\theta$ and $n_g^\theta$ are, respectively, the charging energy and the offset charge corresponding to the $\theta$ coordinate. The flux dependence of the potential energy is given by the coefficients $E_{2}(\varphi_{\text{ext}})=E_{\alpha} - E_{\beta}\cos(\varphi_{\text{ext}})$ and $E_{1}(\varphi_{\text{ext}})=E_{\gamma}\cos(\varphi_{\text{ext}}/2)$, where $E_{\alpha}$, $E_{\beta}$ and $E_{\gamma}$ are constants dependent on the qubit design parameters and studied below.

\begin{figure}
    \centering{}
    \includegraphics[scale=0.81]{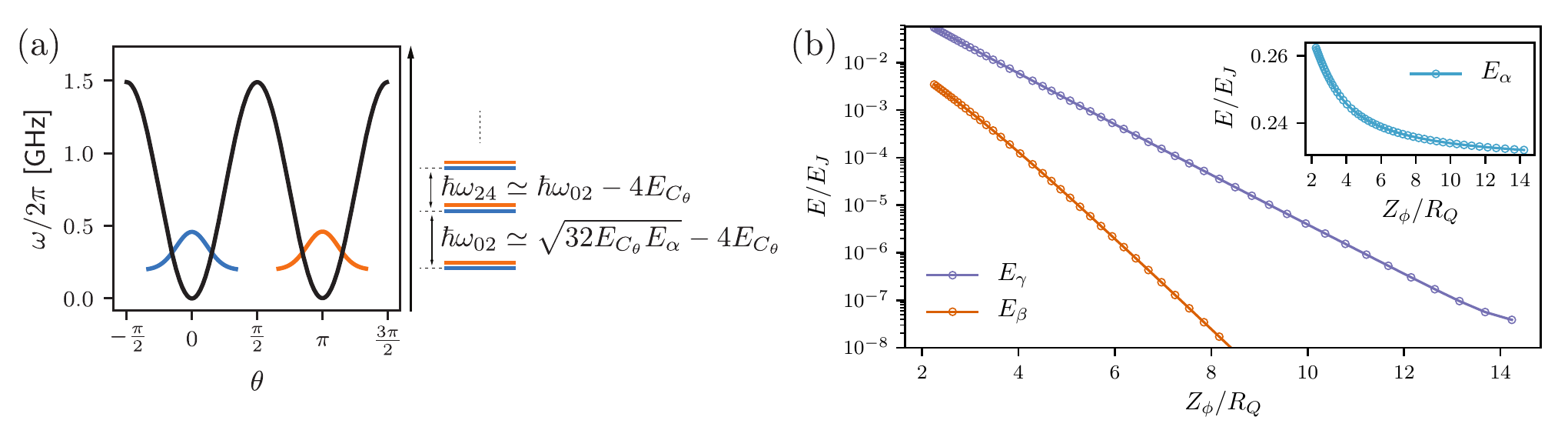}
    \caption{(a) Effective one-dimensional potential (black) and wave functions for the two lowest lying energy states (color), extracted using a Born-Oppenheimer approach (see \cref{app:1d_effective_Hamiltonian}). 
    To the right of the effective potential we show a schematic of the energy diagram (not to scale). In the moderate-to-deep \zp{} regime, the low-energy spectrum consists of nearly degenerate doublets in a weakly anharmonic ladder, closely resembling a transmon qubit spectrum with each transmon level replaced by a doublet. The two lowest doublets are split by approximately $\sqrt{32E_{C_\theta}E_{\alpha}}$, corresponding to the plasma frequency of the $\pi$-periodic Josephson element. (b) Energy parameters of the one-dimensional Hamiltonian \cref{eq:effective_qubit_Hamiltonian} as a function of $Z_{\phi}$ for fixed $Z_{\theta}$. We observe an exponential suppression of both $E_{\beta}$ and $E_{\gamma}$, indicating that the qubit becomes a flux-insensitive $\pi$-periodic Josephson element in the deep \zp{} regime. $E_{\alpha}$ remains almost unchanged in comparison. Circuit parameters: $E_L/\hbar\omega_p\in[1.25\times 10^{-4}, 5\times 10^{-3}]$ and $(E_{C_{\phi}}/\hbar\omega_p,E_{C_{\theta}}/\hbar\omega_p,E_J/\hbar\omega_p) = (0.25, 0.5\times 10^{-3},0.25)$.}
    \label{fig:exciton}
\end{figure}

In the moderate-to-deep \zp{} regime, the relations $E_{\alpha}\gg E_{C_{\theta}}$ and $E_{\alpha}\gg E_\beta, E_{\gamma}$ are satisfied. The effective one-dimensional potential in \cref{eq:effective_qubit_Hamiltonian} is shown in \cref{fig:exciton} (a) for a set of \zp{} circuit parameters. As a function of flux, the two nearly degenerate minima are detuned one with respect to the other, except at $\varphi_{\text{ext}}=\pi$, where the potential becomes perfectly $\pi$-periodic. With $E_2 \gg E_{C_{\theta}}$, tunneling between the two wells is highly suppressed. In the presence of a small, positive $E_{1}$ ($-\pi < \varphi_\text{ext}<\pi$), the lowest energy state is localized in $\theta=0$ and a nearly degenerate first excited state is localized in $\theta=\pi$. At $\varphi_{\text{ext}}=\pi$, the two minima at $\theta=0$ and $\theta=\pi$ are exactly degenerate and the logical wavefunctions become hybridized independently of the circuit design parameters. For $E_1$ smaller than or comparable to the tunneling rate between the potential wells, hybridization can also occur at $\varphi_{\text{ext}}\neq 0$. 

\cref{fig:exciton} (b) shows the values of $\{E_{\alpha},E_{\beta},E_{\gamma}\}$ obtained from a numerical calculation of the coefficients in \cref{eq:effective_qubit_Hamiltonian} as a function of $Z_\phi/R_Q$ for fixed $Z_\theta$ (see \cref{app:1d_effective_Hamiltonian} for details). We observe an exponential suppression of the $\cos\theta$ potential term relative to the $\cos2\theta$ term, justifying the $\pi$-periodicity suggested by the intuitive picture of co-tunneling of Cooper pairs.
 We note that the effective Hamiltonian \cref{eq:effective_qubit_Hamiltonian} in the limit $E_{1}=0$ resembles that of a transmon qubit, with the crucial distinction that the two minima at $\theta=0$ and $\theta=\pi$ are physically distinct. 
We also note, as shown in the inset in~\cref{fig:exciton} (b), that $E_2\simeq E_\alpha \sim E_J$. The condition $E_2 \gg E_{C_\theta}$ thus translates to $E_J \gg E_{C_\theta}$, or equivalently $Z_\theta \ll R_Q$.

Based on this simple picture, the \zp{} qubit approximately reduces to a device with one effective degree of freedom, $\theta$, whose conjugate charge operator, $n_{\theta}$, determines the Cooper pair number difference between the two circuit islands identified in~\cref{fig:nutshell} (b). Since $n_\theta$ changes in units of two, Cooper pair parity is a conserved quantity and an approximate symmetry of the circuit Hamiltonian. We emphasize that the symmetry is approximate, since for finite $Z_\phi$, the $\cos\theta$ term in~\cref{eq:effective_qubit_Hamiltonian} breaks the symmetry.

\subsection{Qualitative explanation of robustness to noise}

Cooper pair-parity conservation partitions the qubit spectrum into doublets with exponentially small charge sensitivity in the ``transmon limit'' $E_J\gg E_{C_{\theta}}$~\cite{PhysRevLett.112.167001,Koch2007a}. 
The $\pi$-periodicity of the Hamiltonian moreover allows us to draw several qualitative conclusions about the qubit's generic properties.
Formally, we define a symmetry operator $U = \exp(-in_\theta\pi)$ which displaces $\theta$ by $\pi$, and note that
\begin{equation}\label{eq:symmetry_condition}
    U H_{0-\pi}^\text{ideal} U\dg = H_{0-\pi}^\text{ideal} + \dots,
\end{equation}
where the ellipses refer to exponentially small corrections in the deep \zp{} regime, as we have verified above. Denoting the ground state of the Hamiltonian by $\ket 0$ with energy $E_0$, it follows that a second eigenstate with energy exponentially close to $E_0$ is given approximately by
$U \ket 0$. This follows from
$H_{0-\pi}^\text{ideal} \left(U \ket 0\right) = \left(U H_{0-\pi}^\text{ideal} U\dg\right) \left(U \ket 0\right) + \dots = E_0 \left(U\ket 0\right) + \dots$ We can denote this eigenstate by $\ket 1$. Moreover, the argument continues to hold in the presence of any perturbation to the Hamiltonian that respects the (approximate) symmetry~\cref{eq:symmetry_condition},
\emph{i.e.}
\begin{equation}\label{eq:nodephasing}
    \bra 0 V \ket 0 = 
    \bra 1 V \ket 1 + \dots
\end{equation}
where $V$ satisfies $U V U\dg = V + \dots$
It follows that dephasing noise is expected to be exponentially suppressed for \emph{symmetry-preserving} noise processes. In particular, \cref{eq:effective_qubit_Hamiltonian} shows that external flux noise does not break the $\pi$-periodicity [recall that $E_1(\varphi_\text{ext})$ is exponentially suppressed in the deep \zp{} regime].

The condition $E_J\gg E_{C_\theta}$ (or equivalently $Z_\theta \ll R_Q$) moreover leads to exponential suppression of tunneling between the two potential wells located at $\theta = 0$ and $\theta = \pi$, as already discussed.
When the two nearly degenerate ground states are localized in the two different wells, this thus leads to an exponential suppression of bit-flips~\footnote{Depending on the circuit parameters and external flux, the two ground states can be localized in the two different wells, or in some cases symmetric and anti-symmetric superpositions of such localized states~\cite{Dempster2014a}. In the latter case, the $Z$ and $X$ basis are exchanged.}
\begin{equation}\label{eq:nobitflips}
\bra 0 V \ket 1 = 0 + \dots,
\end{equation}
for $V$ any weak perturbation to the Hamiltonian that is local in phase space, \emph{i.e.}, any low-degree polynomials in $\{\phi,\theta,q_\phi,q_\theta\}$. \cref{eq:nodephasing,eq:nobitflips} lead together to the remarkably long coherence times expected for the qubit in the deep \zp{} regime, as recently confirmed quantitatively in Ref.~\cite{groszkowski2017coherence}.

\section{Coupling to external circuitry}
\label{sec:coupling}

\subsection{General remarks about coupling strategies}

With the goal of controlling and measuring the \zp{} qubit, we now outline different strategies to couple the qubit to external degrees of freedom.
Noise protection in the \zp{} qubit is achieved at a high price: The protection from bit-flips implies negligible matrix elements for qubit transitions making many coupling schemes inefficient. Moreover, great care has to be taken to not introduce coupling circuitry explicitly breaking the $\pi$-periodicity, opening the qubit to dephasing noise.
Some general remarks about coupling strategies can be made based on the qualitative discussion of the \zp{} qubit in the previous section.

\subsubsection{Direct inductive coupling}
Any galvanic linear inductive coupling to the four circuit nodes leads to contributions of the generic form $\sim E_{L,\theta}\theta^2$ to the Hamiltonian, explicitly breaking the \zp{} periodicity and lifting the groundspace degeneracy. It might be possible to approximately restore the \zp{} periodicity by using superinductors such that $E_{L,\theta}\to 0$. However, this in turn leads to negligible coupling to any external circuitry, rendering such an approach ineffective.

\subsubsection{Mutual inductive coupling}
As mentioned above, in the limit $L\to\infty$, moderate variations of the external flux through the qubit loop do not break the \zp{} symmetry, such that mutual inductive coupling can potentially be a symmetry preserving coupling mechanism. However, for precisely the same reason that the qubit is highly insensitive to flux noise~\cite{groszkowski2017coherence}, control and readout strategies based on mutual inductive coupling are ineffective. Large external flux excursions, in contrast, can be used to move between regimes where the logical states are localized in different potential wells, to a regime where they are in a superposition of both wells. We discuss exploiting this in a control strategy in~\cref{sec:QuantumNOTGate}.

\subsubsection{Capacitive coupling}
Capacitive coupling to the circuit nodes has the advantage that it only couples directly to the charge degrees of freedom, leaving the \zp{} periodicity and the two-island topology in~\cref{fig:nutshell} (b) intact. Moreover, as long as the coupling capacitances are kept small, they should not compromise the inequality~\cref{eq:0pi_regime}. In general, the extremely small matrix elements coupling the logical qubit states make many conventional control and readout strategies inefficient. Nevertheless, we show below that capacitive coupling can be used to perform device spectroscopy in a moderate-to-deep regime of parameters, enable single-qubit control by means of fast voltage drives, and cool the parasitic $\zeta$-mode to improve the qubit coherence times. 

\subsubsection{Non-linear symmetry-preserving inductive coupling}
Although we have argued that any straightforward coupling strategy based on inductive elements is either ineffective or breaks the qubit's protection from noise, it might still be possible to engineer \emph{non-linear} inductive couplers that respects the \zp{} symmetry. This means that the inductive contribution to the energy has to satisfy $E_\text{coupler}(\theta) = E_\text{coupler}(\theta + \pi) + \dots$, where the ellipses again refer to terms that vanish in the deep \zp{} regime. In Ref.~\cite{Brooks2013}, it was proposed that such a coupling mechanism can be achieved by using a tunable Josephson coupler (SQUID loop) connecting the \zp{} qubit to an LC oscillator. For the qubit to remain protected, the LC oscillator with impedance $Z_r$ is required to satisfy $Z_r \gg R_Q$, much like the internal $\phi$ mode of the \zp{} circuit. Moreover, it was shown that such a coupling could be used to enact one- and two-qubit phase gates. We briefly return to this scheme below, and point out some additional challenges which have previously been overlooked. 

\subsection{Addressing the \zp{} qubit degree of freedom}
\label{sec:CouplingStrategies}
 
\begin{figure}
    \centering{}
    \includegraphics[scale=0.6]{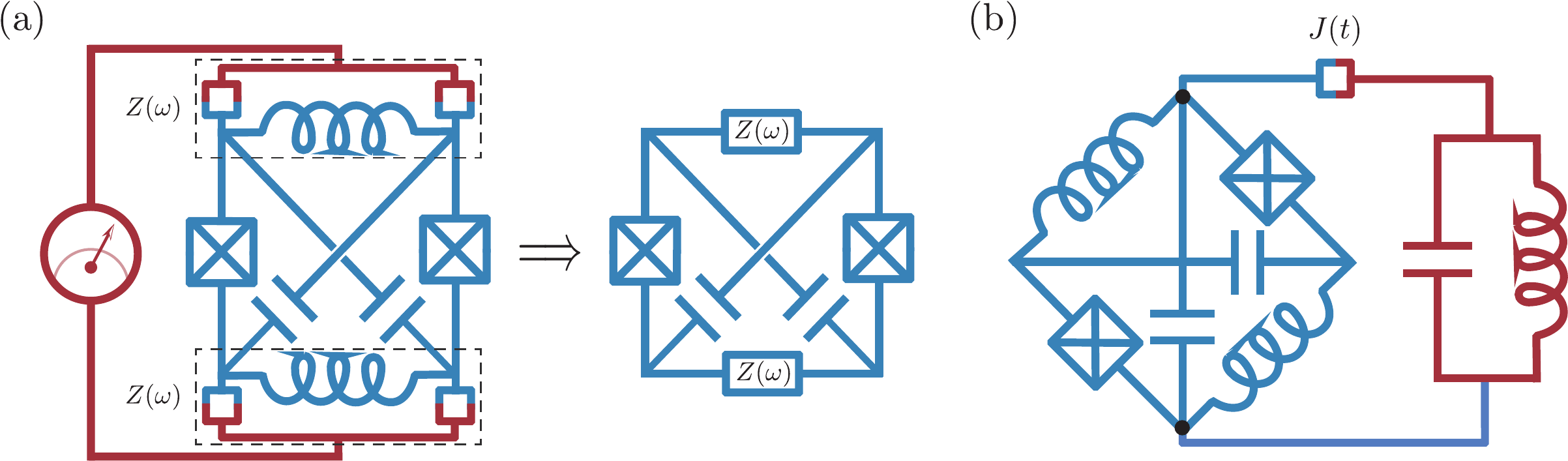}
    \caption{(a) Addressing the qubit main degree of freedom $\theta$. The small box-shaped couplers are used to represent arbitrary coupling circuit elements, and the meter represents a readout or control circuit. As illustrated by a dashed frame, the required coupling circuitry (in red) necessarily shunts the device superinductors, leading to a combined link impedance $Z(\omega)$ that can compromise the qubit operation. (b) Coupling layout originally considered in \cite{Brooks2013}. The interaction strength $J(t)$ stands for the potential energy of a flux-tunable device.}
    \label{fig:coupling_theta}
\end{figure}

Coupling to the qubit mode $\theta$ in the \zp{} circuit has an additional challenge, beyond the general points already made above.
Because the coordinate $\theta$ is a combination of phase operators of all nodes of the circuit [see \cref{eq:normal_modes}], addressing only this coordinate requires a coupling element acting symmetrically on both ports of each superinductor.  As illustrated schematically in \cref{fig:coupling_theta} (a), where the boxes represent unspecified coupling elements and could be capacitive or inductive in general, this coupling circuitry necessarily shunts the \zp{} qubit superinductors. According to the discussion in \cref{sssec:QubitStructure}, if the impedance of the coupler is not greater than or comparable to that of the qubit superinductors, this effect can potentially compromise regime of operation of the device. At first glance, a possible solution to this problem appears to be the use of additional superinductors replacing each of the box-shaped couplers in \cref{fig:coupling_theta} (a). However, this would lead to an inductive shunt of the \zp{} circuit islands identified in \cref{fig:nutshell} (b) through the readout or control circuit [represented by a meter in \cref{fig:coupling_theta} (a)], breaking the Cooper pair parity symmetry. 

An alternative coupling scheme considered in Ref.~\cite{Brooks2013} is illustrated in~\cref{fig:coupling_theta} (b). In this scheme, the coupling element (\emph{i.e.} the box in the figure) is a SQUID loop giving a tunable Josephson element between the \zp{} circuit and the LC oscillator. This coupling layout overcomes the difficulty described at the beginning of this section by relaxing the symmetry requirements of the coupling circuitry. However, this leads to an interaction Hamiltonian that involves both $\theta$ and the spurious $\zeta$-mode 
\begin{equation}
U_J = -J(t)\cos(\theta+\zeta-\phi_r),
\label{eq:potential_energy_J}
\end{equation}
where $J(t)$ is the tunable Josephson energy of the coupling element, and $\phi_r$ the resonator phase operator. We have previously shown that the $\zeta$-mode frequency goes to zero in the deep \zp{} regime leading to diverging thermal occupation of this mode~\cite{groszkowski2017coherence}, something which was not taken into account in Ref.~\cite{Brooks2013}. The impact of thermal fluctuations due to the \zp{} circuit internal modes thus requires further study and we propose in \cref{sec:CoolingTheZetaMode} a cooling scheme that can help approximate the ideal behavior considered in \cite{Brooks2013}.

Based on this discussion, the most viable option for near-term experiments appears to be the use of capacitive coupling. This means replacing the box-shaped couplers in \cref{fig:coupling_theta} (a) by capacitors. Formally, small coupling capacitors operate as high-impedance links while preserving the circuit islands. The coupling capacitances must be kept small to ensure $Z_\phi/R_Q \gg 1$ since these add to $C_{\phi}$ [see \cref{eq:mode_capacitances}]. Therefore, a downside of this approach resides in the fact that the capacitive couplings cannot be very large. Nevertheless, we find that capacitive coupling allows for significant dispersive shifts (\cref{sec:Dispersive}), and a fast, single-qubit gate (\cref{sec:QuantumNOTGate}) .

We emphasize that the control strategies we consider in the following are not fault-tolerant in the sense of~\cite{kitaev2006protected,Brooks2013}. They either rely on operating the qubit in a regime that is not fully protected, or involve populating higher and less robust excited states. Nevertheless, these operations can be performed with high fidelity and are thus suitable for implementation in realistic devices in the near future.

\subsection{Capacitive coupling to voltage sources}
\label{sssec:VoltageDriveCoupling}

\begin{figure}
    \centering{}
    \includegraphics[scale=1.19]{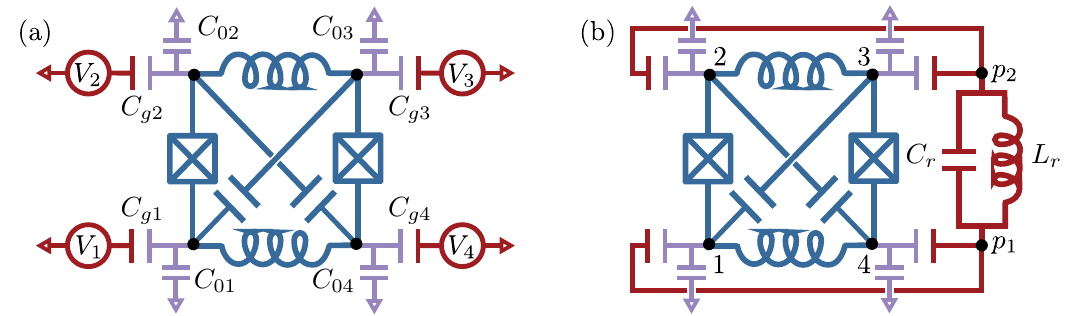}
    \caption{
    (a) Lumped element model for the \zp{} circuit coupled to microwave voltage sources, including gate and ground capacitances for each circuit node. (b) \zp{} circuit connected to a resonator with nodes $p_1$ and $p_2.$ The shown coupling layout couples the resonator charge operator to $q_{\theta},$ and corresponds to the second row in \cref{table:coulping_the_0-pi_to_a_resonator}.}
    \label{fig:circuit_diagram}
\end{figure}

We now consider the \zp{} circuit in the presence of voltage sources $V_i$ connected to the nodes $i=1,...,4$ of the circuit, as shown in \cref{fig:circuit_diagram} (a).
Since we have found that circuit element disorder is a limiting factor for the qubit coherence for parameters within reach of current experiments~\cite{groszkowski2017coherence}, we include such effects here. In particular, we account for any superinductance and Josephson energy asymmetries, denoted by $dE_L$ and $dE_J,$ respectively, as well as capacitance asymmetries, denoted by $dC_J$ and $dC$. Additionally, there can be disorder in the gate capacitances ($dC_{g_i}$), as well as in the parasitic capacitances to ground ($dC_{0_i}$), such that the node gate and ground capacitances for node $i$ are $C_{g_i} = C_g (1+ dC_{g_i})$ and $C_{0_i} = C_0 (1+ dC_{0_i}),$ respectively. We note that, in practice, the stray capacitances may arise from the superinductances and the large capacitors of the \zp{} circuit \cite{Masluk2012}. Following the standard approach to circuit quantization \cite{devoret1995quantum,burkard2004multilevel} we find the Hamiltonian
\begin{equation}
H = H_{0-\pi} + H_{\text{drive}}^\text{symm}  + H_\text{drive}^\text{asymm},
\label{eq:qubit_hamiltonian}
\end{equation}
where $H_{0-\pi} = H_{0-\pi}^\text{symm} + H_{0-\pi}^\text{asymm}$ describes the un-driven qubit. The first contribution
\begin{equation}
    H_{0-\pi}^\text{symm} = \sum_{\mu}\frac{q_{\mu}^2}{2C_{\mu}} - 2E_J\cos\theta\cos\qty(\phi-\varphi_{\text{ext}}/2) + E_L (\phi^2 + \zeta^2),
    \label{eq:0pi_hamiltonian}
\end{equation}
with $q_{\mu}/2e=-i \partial_{\mu}$ for $\mu=(\phi,\theta,\zeta,\Sigma)$
is the ideal \zp{} Hamiltonian, where we now explicitly include the $\zeta$ and $\Sigma$ degrees of freedom and the mode capacitances $C_\mu$ defined explicitly in~\cref{app:gate_and_ground_assymetries}.
On the other hand,
\begin{equation}
    \label{eq:Hcoupl}
    H_{0-\pi}^\text{asymm} = -\frac{C\,dC}{C_{\zeta} C_{\theta}}q_{\theta}q_{\zeta} -\frac{C_J\,dC_J}{C_{\phi} C_{\theta}}q_{\phi}q_{\theta} 
    + E_J dE_J \sin\theta\sin\qty(\phi-\varphi_{\text{ext}}/2) + E_L dE_L \phi \zeta
    + H_{dC_g,dC_0},
\end{equation}
describe unwanted spurious couplings between the circuit modes to leading order in circuit element disorder.
The last term $H_{dC_g,dC_0}$ is a purely capacitive term accounting for disorder of the gate and ground capacitances, and its full expression can be found in \cref{app:gate_and_ground_assymetries}. Since these capacitances are expected to be much smaller than the internal circuit capacitances $C$, we however neglect $H_{dC_g,dC_0}$ in the remainder of this work.
Finally, the drive term
\begin{equation}
    H^\text{symm}_\text{drive} = \sum_{\mu} \frac{C_g}{C_{\mu}} V_{\mu}q_{\mu},
\end{equation}
describe voltage drives of the four normal modes where $V_\mu$ is defined in terms of the external node voltages $V_i$ with $i=1,\dots,4$ according to the transformation rule in~\cref{eq:normal_modes}. Circuit element disorder furthermore introduces additional drive terms. This is accounted for by the Hamiltonian $H^\text{asymm}_\text{drive}$, given explicitly in~\cref{app:gate_and_ground_assymetries}.

As can be seen from~\cref{eq:Hcoupl} the coupling between the qubit degrees of freedom $\{\phi,\theta\}$ and the spurious $\zeta$-mode appears when the large circuit capacitors or the superinductors are not symmetrical: $dC\neq 0$ or $dE_L\neq0,$ respectively. As we have shown recently \cite{groszkowski2017coherence}, this leads to the limiting contribution to the qubit's coherence time for realistic parameters due to photon shot noise for the $\zeta$-mode. We return to how to alleviate this issue in~\cref{sec:CoolingTheZetaMode}.

\subsection{Capacitive coupling the \zp{} qubit to a microwave resonator}

\label{sssec:CapacitivelyCouplingToResonator}
With the goal of controlling and reading out the \zp{} qubit, we consider its capacitive coupling to a microwave resonator as illustrated in~\cref{fig:circuit_diagram} (b). The Hamiltonian of the combined qubit-resonator system can be obtained from \cref{eq:qubit_hamiltonian}, by adding the free resonator Hamiltonian, $H_r = \hbar\omega_r a_r^\dagger a_r$, and letting $V_{\mu}$ correspond to the resonator voltage \footnote{Although applicable in most circuit QED setups, a cautionary remark is that this procedure is only valid in the weak capacitive coupling regime, where the capacitances of coupled qubit and resonator modes are large compared to the coupling capacitance.}. \cref{table:coulping_the_0-pi_to_a_resonator} specifies the replacement rules for the voltages $V_{\mu}$ in \cref{eq:qubit_hamiltonian} that produce the qubit-resonator interaction Hamiltonian. Three possible coupling layouts addressing the \zp{} degrees of freedom $\theta$ [shown in \cref{fig:circuit_diagram} (b)], $\phi$ and $\zeta$ are considered. These capacitive coupling schemes are employed in~\cref{sec:Dispersive} for dispersive readout strategies, in \cref{sec:QuantumNOTGate} to drive qubit transitions via multiple excited levels, and in \cref{sec:CoolingTheZetaMode} to cool the low-frequency $\zeta$-mode as a strategy to enhance the qubit coherence times.
\begin{table}
\centering
\begin{tabular}{cccc}\hline\hline
 {\zp{} mode}  & {\zp{} nodes connected to $p_1$}  & {\zp{} nodes connected to $p_2$}  & {Replacement rule in \cref{eq:qubit_hamiltonian}} \\\hline\hline
  $\phi$  &$1,3$     &$2,4$   &$V_{\mu}\rightarrow \delta_{\mu,\phi}\,V_r$  \\
  $\theta$  &$1,4$     &$2,3$   &$V_{\mu}\rightarrow \delta_{\mu,\theta}\,V_r$  \\
  $\zeta$  &$3,4$     &$1,2$   &$V_{\mu}\rightarrow \delta_{\mu,\zeta}\,V_r$  \\\hline\hline
\end{tabular}
\caption{Capacitively coupling the \zp{} qubit to an external resonator. The second and third columns specify which \zp{} nodes are connected to the resonator nodes $p_1$ and $p_2,$ respectively, thus determining the replacement rule for $V_{\mu}$ in \cref{eq:qubit_hamiltonian}, as indicated in the fourth column ($\delta_{\mu,\nu}$ is here the Kronecker delta). The resonator voltage is given by
$V_r = q_r/C_r = i V_{\text{rms}} (a_r^{\dag} - a_r)/2$, where $V_{\text{rms}}$ is the resonator root-mean-squared voltage fluctuations in the ground state, and $a_r$ the resonator annihilation operator.}
\label{table:coulping_the_0-pi_to_a_resonator}
\end{table}

\section{Dispersive readout}
\label{sec:Dispersive}

The transmon-like structure of the \zp{} energy spectrum illustrated in~\cref{fig:nutshell} (d) suggests that we might exploit known techniques for dispersive readout and control for transmon qubits~\cite{Koch2007a}.
The strong symmetry between the two potential wells at $\theta=0$ and $\theta=\pi$, however, means that each ``transmon level'' is split into a doublet, leading to important differences in dispersive coupling for a \zp{} qubit as compared to a conventional transmon.

 Dispersive coupling to a resonator relies on having unequal qubit-dependent dispersive shifts of the resonator frequency for the two logical states $\ket 0$ and $\ket 1$. We compute the dispersive shifts numerically assuming capacitive coupling between either of the two \zp{} modes $\{\theta,\phi\}$ and a readout resonator of frequency $\omega_r/2\pi$ (see \cref{table:coulping_the_0-pi_to_a_resonator}). Denoting by $a_r$ the annihilation operator of the readout resonator and including $M$ qubit levels, the qubit-resonator Hamiltonian can be written as
\begin{equation}
\begin{aligned}
H ={}& \sum_{i=0}^M \hbar \omega_i \sigma_{ii} + \hbar\omega_r a_r^{\dag} a_r  + \sum_{i,j=0}^M g^{\mu}_{ij}\sigma_{ij}(a^{\dag}_r + a_r),
\end{aligned}
\label{eq:Hamiltonian_for_Dispersive}
\end{equation} 
where $\sigma_{ij} = \ket i \bra j$, $g^{\mu}_{ij} = \frac{C_g}{C_{\mu}}(eV_{\text{rms}})\bra{i}n_{\mu}\ket{j}$ and $\mu=\{\theta, \phi\}$. Note that the resonator drive has not been explicitly included in \cref{eq:Hamiltonian_for_Dispersive}. In the dispersive regime defined by $|\Delta_{ij}|\gg|g_{ij}|\sqrt{\bar{n} + 1}$, where $\Delta_{ij} = (\omega_{i} - \omega_j) - \omega_r$ and $\bar{n}$ is the mean number of photons in the resonator, the above Hamiltonian takes the form \cite{zhu2013circuit}
\begin{equation}
\begin{aligned}
H ={}& \sum_i^M \hbar (\omega_i + \Lambda^{\mu}_i) \sigma_{ii} + \hbar\omega_r a_r^{\dag} a_r + \sum_i^M \hbar\chi^{\mu}_i \sigma_{ii}a_r^{\dag}a_r\\
\simeq{}& \frac{\hbar \tilde\omega_q}{2}\sigma_z + \hbar\tilde\omega_r a_r\dg a_r + \hbar\chi^\mu \sigma_z a_r\dg a_r,
\end{aligned}
\label{eq:Hamiltonian_Dispersive}
\end{equation} 
where the dispersive shift of the $i^{\text{th}}$ qubit level is given by $\chi^{\mu}_i = \sum_j^M (\chi^{\mu}_{ij} - \chi^{\mu}_{ji})$, with $\chi^{\mu}_{ij} = |g^{\mu}_{ij}|^2/\Delta_{ij}$, and $\Lambda^{\mu}_i =\sum_j^M \chi^{\mu}_{ij}$ is the corresponding Lamb-shift. The second line in~\cref{eq:Hamiltonian_Dispersive} is a two-level truncation where we have defined
$\sigma_z = \ket 1\bra 1 - \ket 0 \bra 0$,
$\tilde\omega_q = \omega_1 + \Lambda_1^\mu - \omega_0 - \Lambda_0^\mu$,
$\tilde\omega_r = \omega_r + \chi_0^\mu/2 + \chi_1^\mu/2$
and $\chi^\mu = (\chi_1^\mu-\chi_0^\mu)/2$.

We investigate the dispersive coupling $\chi^{\mu}$ as a function of the \zp{} design parameters. We choose the resonator frequency such that $\chi^{\mu}$ is maximized while ensuring the validity of the dispersive approximation. For the case of coupling to $\theta$, we observe that $\chi^{\theta}$ is heavily attenuated in the parameter space corresponding to a moderate-to-deep \zp{} qubit regime. This is due to the fact that, in contrast to a transmon qubit, the strong symmetry between the left and right potential wells of the \zp{} qubit leads to vanishing dispersive coupling to the resonator for most parameters. Moreover, since the external flux does not break this symmetry [see \cref{eq:effective_qubit_Hamiltonian}], $\chi^{\theta}$ can only slightly change by flux excursions. 

\begin{figure}
    \centering{}
    \includegraphics[scale=1.2]{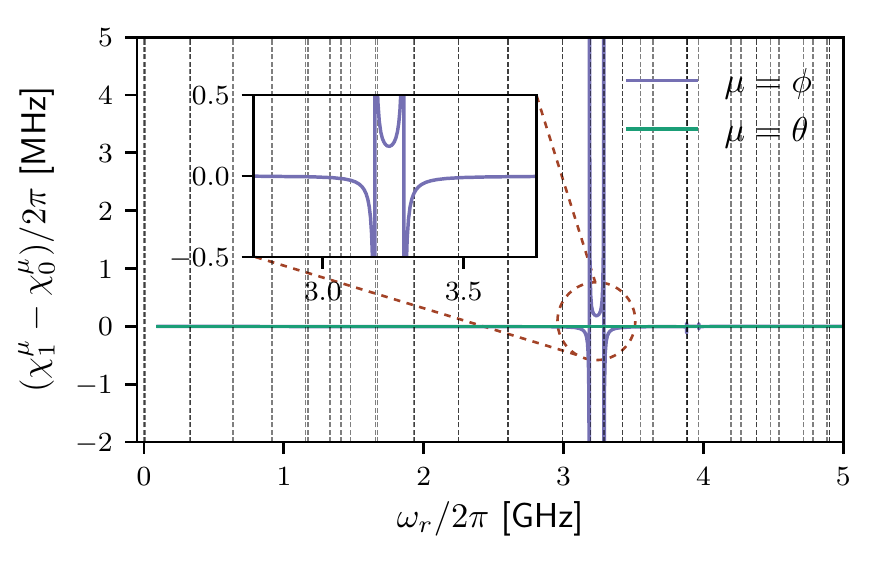}
    \caption{Dispersive shift for the groundstate doublet of the \zp{} qubit as a function of the readout resonator frequency $\omega_r/2\pi$. The qubit spectrum is shown in black dashed lines. Note that many of such lines are superimposed due to the doublet structure of the qubit spectrum, and in particular for the groundstate doublet around 0 GHz. For $\phi$ coupling, we observe a remarkable increase of the dispersive shift in the highlighted region, reminiscent of the straddling regime of a transmon qubit. Here the qubit design parameters correspond to a moderate-to-deep \zp{} regime at $\varphi_{\text{ext}}=0$, with $(E_L/\hbar\omega_p,E_{C_{\phi}}/\hbar\omega_p,E_{C_{\theta}}/\hbar\omega_p,E_J/\hbar\omega_p) = (1.25\times10^{-3},0.374, 1.25\times 10^{-4},0.167)$. Furthermore, we assume $C_g/C_{\mu}=0.2$.}
    \label{fig:straddling}
\end{figure}

Quite surprisingly, however, we find a significant dispersive shift for the coupling operator $n_{\phi}$, as shown in \cref{fig:straddling}. This behavior is qualitatively reminiscent to what is known as the straddling regime for the transmon qubit, in which the dispersive shift can increase by orders of magnitude \cite{Koch2007a}. Note, however, that the narrow straddling-like regime indicated in~\cref{fig:straddling} is related to the splitting of doublets rather than the plasma frequency separation between two sets of doublets, and the large number of qubit levels involved makes the situation more complex than in a transmon.
Interestingly, the value of $\chi^{\phi}$ adds a significant contribution from qubit levels generated by excitations of the $\phi$ degree of freedom, which are not captured by the effective model in \cref{sssec:QubitStructure}.

In practice, we find that the absolute value of $\chi^{\phi}/2\pi$ does not increase beyond a few hundred kHz in a moderate-to-deep \zp{} parameter regime. This would lead to rather slow readout and resonator mediated gates as compared to those for the transmon qubit \cite{Cross2015,Puri2016}. However, an appreciable $\chi^{\phi}$ could be useful to resolve the qubit nearly-degenerate doublet by means of spectroscopy, and thus play an important role for device characterization. Moreover, we emphasize that the example parameter set in~\cref{fig:straddling} is rather deep in the \zp{} regime, where qubit lifetimes are predicted to be extremely long~\cite{groszkowski2017coherence}. Reduced gate and and readout times might therefore be an acceptable compromise.

\section{Single-qubit control through multilevel excursions}
\label{sec:QuantumNOTGate}

\subsection{Qualitative picture}
\label{sssec:IntroductionGate}

In this section, we study a process achieving population inversion between the logical qubit states. Such an operation seems challenging at first, given that, by design, the off-diagonal matrix elements of charge and phase operators in the qubit subspace are exponentially small in the deep \zp{} regime~\cite{Dempster2014a,groszkowski2017coherence}. In particular, transition matrix elements for the charge operator can easily be $10^{-8}$ times smaller than those for the transmon qubit. We overcome this situation by exploiting the multilevel structure of the device for gate operations.

A first possible approach to circumvent the small overlap between logical states relies on Raman transitions, with the advantage of only virtually populating states outside of the protected subspace. However, in \cref{app:RamanGates}, we show that due to destructive interference the amplitudes of Raman processes in general vanish as the system approaches the deep \zp{} limit. For this reason, we consider instead a gate scheme that temporarily populates excited states during the gate~\cite{douccot2012physical}. The gate lifts some of the qubit's protection from noise, as it populates higher energy levels. Nevertheless, the proposed strategy requires leaving the qubit subspace only for very short times, and we consequently find high fidelities for a broad range of parameters.

\begin{figure}
    \centering{}
    \includegraphics[scale=1.0]{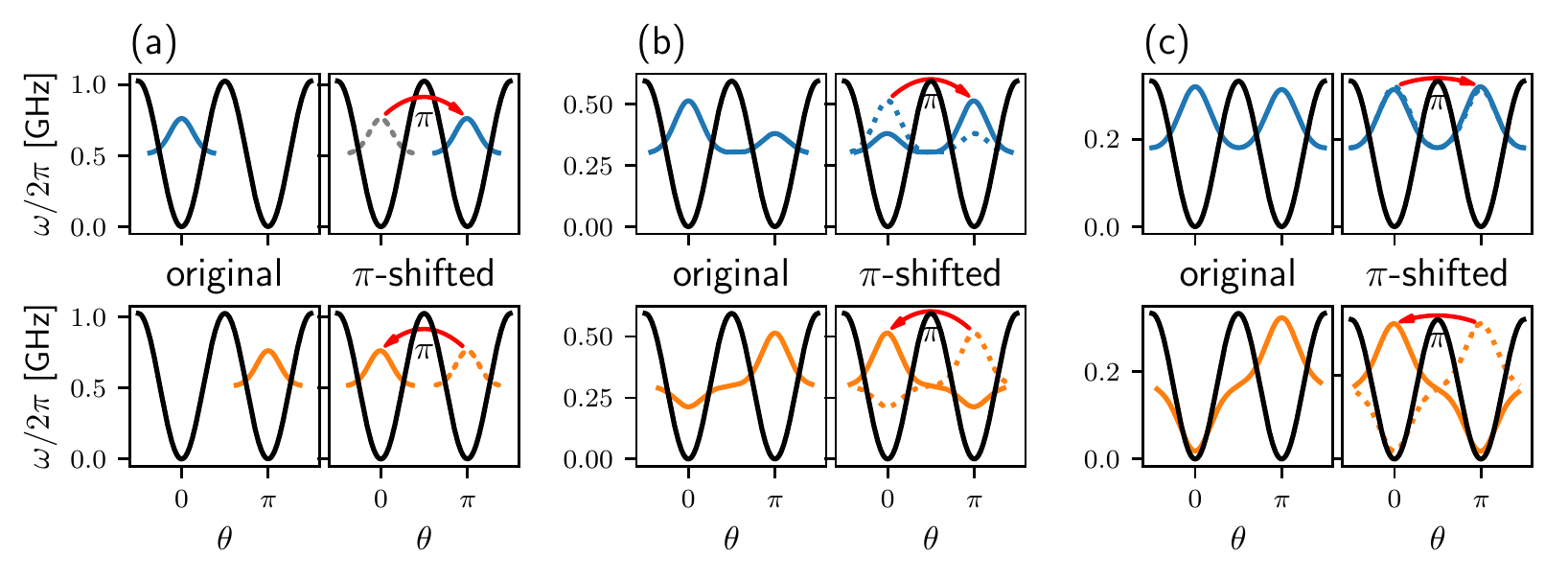}
    \caption{Single-qubit gate operation within the effective \zp{} model for three chosen configurations: (a), (b) and (c) correspond (respectively from top to bottom) to the qubit parameters highlighted with light-blue dots in \cref{fig:fidelity_plots} (b). Ground (in blue) and excited (in orange) wave functions are displayed on the top- and bottom- left corner of each panel, respectively. To the right of the panels, we show the effect of a $\pi$-shift on such wave functions, demonstrating the gate operation. Note that because of the $2\pi-$periodicity of the \zp{} potential (in black), a $\pi$-shift to the left is equivalent to a a $\pi$-shift to the right. The gate implements a Pauli $X$ operation for the case (a), a Hadamard for (b), and a Pauli $Z$ for (c).}
    \label{fig:wave_functions_and_gate}
\end{figure}

An intuitive understanding of the proposed gate can be gained by returning to the effective one-dimensional model for the \zp{} qubit presented in~\cref{sssec:QubitStructure}. In this simplified scenario, we have already suggested that logical $\ket{1}$ can approximately be obtained from $\ket{0}$ using a displacement by $\pi$ along $\theta.$ Such an operation corresponds to the unitary $\exp(-in_{\theta}\pi)$, which can be generated by voltage driving the qubit.
The precise logical action of such a displacement, however, depends on circuit and external parameters, as this determines the structure of the logical wave functions in the two wells.
\cref{fig:wave_functions_and_gate} shows the logical wave functions corresponding to three different points in parameter space that will be studied in detail below. The figure shows the logical wave functions before and after a shift of $\theta\to\theta + \pi$ that represents the gate operation. When ground and excited states are respectively localized in the $\theta=0$ and $\theta=\pi$ wells of the \zp{} qubit potential [\cref{fig:wave_functions_and_gate} (a)], a $\pi$-shift corresponds to a Pauli $X$ operation. If $E_J/E_{C_\theta}$ is lowered, the hybridization of the qubit logical states increases. In the situation illustrated in \cref{fig:wave_functions_and_gate} (b), the logical wave functions are no longer perfectly localized and the gate implements a Hadamard operation. If, instead, the logical wave functions are completely hybridized [\cref{fig:wave_functions_and_gate} (c)], a $\pi$-shift corresponds to a Pauli $Z$ gate. We can alternatively achieve the same wave function control by varying the external flux where $\varphi_\text{ext}=0$ corresponds to localized wave functions (for large $E_J/E_{C_\theta}$) and $\varphi_\text{ext}=\pi$ to completely hybridized wave functions.
In the following sections we study this qualitative picture in detail.

\subsection{Gate fidelity with respect to circuit parameters}
\label{sssec:QuantumNOTGateFidelityParams}

When the full \zp{} Hamiltonian is considered, the asymmetry of the two-dimensional logical wave functions along the $\phi$-direction [see~\cref{fig:1D_model} (c)] make clear that $\ket{1}$ and $\ket{0}$ cannot be simply exchanged by means of a $\theta$-translation alone. Taking this into consideration, this section studies the gate employing the full circuit Hamiltonian \cref{eq:qubit_hamiltonian}. We characterize the gate fidelity as a function of the \zp{} design parameters, and analyze the effect of circuit element disorder and pulse shaping in the following sections.

We first consider a square microwave voltage pulse applied to the qubit and driving the $\theta$ coordinate, in absence of circuit element disorder. In \cref{eq:qubit_hamiltonian}, this situation corresponds to setting all $V_{\mu}$ to zero with the exception of $V_{\theta}$, such that the circuit Hamiltonian reads 
\begin{equation}
H = H_{0-\pi}^\text{symm} + \frac{C_g}{C_{\theta}} V_{\theta}(t)q_{\theta}.
\label{eq:qubit_hamiltonian_theta_drive}
\end{equation}
In this section, we assume that the microwave drive is turned on at $t=0$, reaching an amplitude $V_{\text{sq}}$ for a period of time $t_g$. The effect of pulse shaping is analyzed below. To determine the optimal drive strength given the \zp{} design parameters, we compute the multilevel evolution operator as a function of the pulse parameters $(V_{\text{sq}},t_g)$, and minimize its distance to a unitary acting only on the qubit subspace \footnote{Denoting $u_{\text{red}}$ as the reduced propagator, we define its distance to a unitary as $d(u_{\text{red}}) = \max_s|1-s|,$ where $\{s\}$ are the singular values obtained from the singular value decomposition $u_{\text{red}}=W_{\text{pre}}SW^{\dag}_{\text{post}}.$ The closest unitary is defined as $u_{\text{closest}}=W_{\text{pre}}W^{\dag}_{\text{post}},$ and identifies the qubit rotation that the voltage drive implements on the logical subspace \cite{goerz2015optimizing,danielreich}.}. This procedure ensures that leakage errors are kept as small as possible at the end of the gate. For the optimal drive configuration, we determine the closest qubit unitary to the multilevel propagator, and compute the average gate fidelity of the latter with respect to the former, including leakage errors \cite{johansson2012qutip}.

\begin{figure}
    \centering{}
    \includegraphics[scale=1.0]{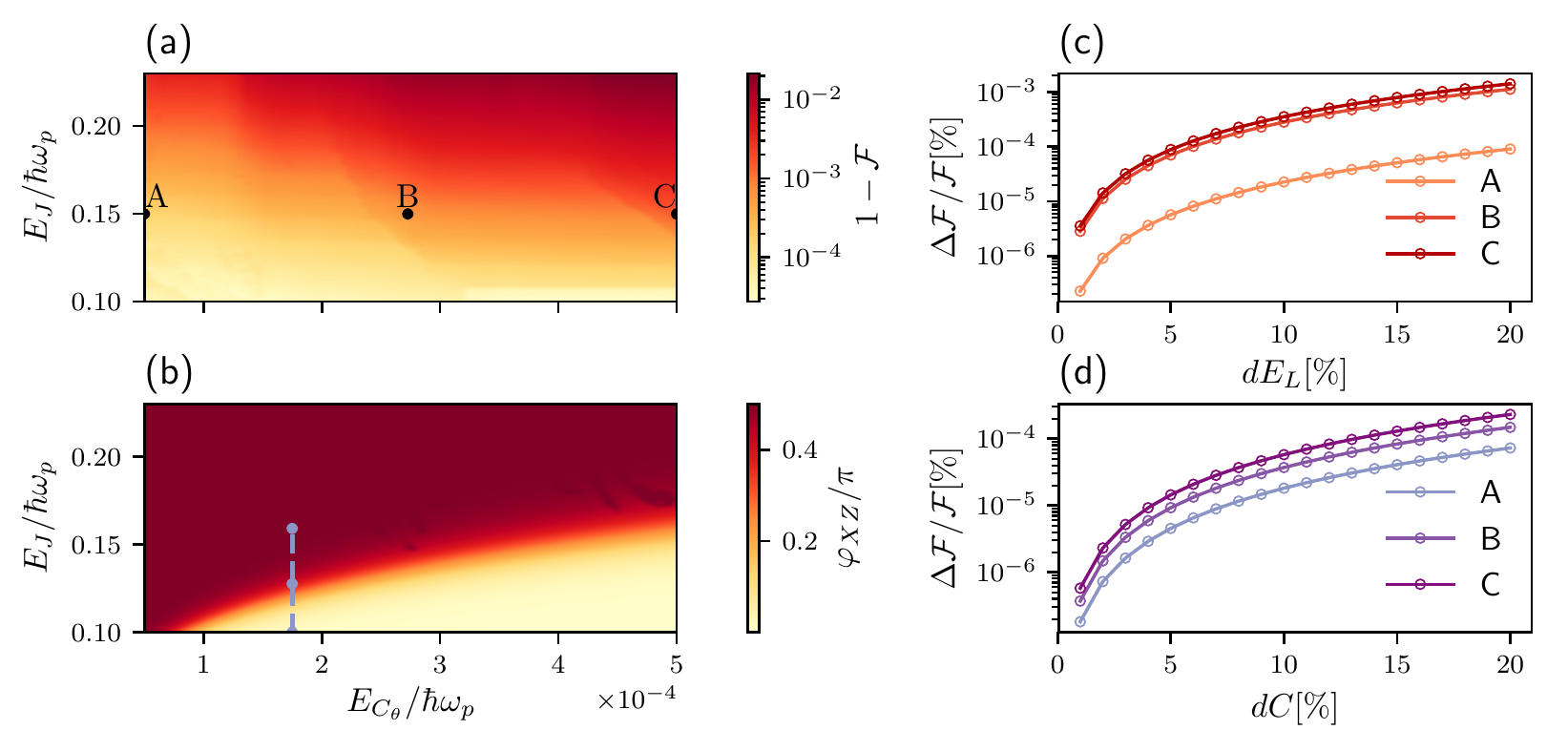}
    \caption{Single-qubit gate infidelity and logical action on the Bloch sphere. (a) Gate infidelity for a device with no disorder, computed from the unitary (non-dissipative) dynamics of the system. As $E_{C_{\theta}}/\hbar\omega_p$ and $E_J/\hbar\omega_p$ increase, we observe a decrease in gate fidelity. This is qualitatively understood as the effect of increasingly longer multilevel excursions during the gate. (b) Induced rotation on the Bloch sphere. Here we show the polar angle $\varphi_{\text{XZ}}$ for the $XZ$ plane, while the azimuthal angle $\varphi_{\text{XY}}$ remains bounded below $10^{-5}$. We note that as $E_J/\hbar\omega_p$ is reduced, the qubit rotation smoothly interpolates between a Pauli $X$ and Pauli $Z$. The light-blue dots are used as a reference for \cref{fig:wave_functions_and_gate}. Panels (c) and (d) show the relative change in the gate fidelity for the configurations A, B and C in panel (a), when dissipation and disorder in $E_L$ and $C$ are included. The gate fidelity proves to be robust to parasitic coupling to the $\zeta$-mode for moderate amounts of circuit element disorder, and it is not significantly affected by dissipation. The latter is a consequence of the fast gate time compared to the expected decoherence rates. For numerical reasons, simulations in panel (c) and (d) assume a cooled $\zeta$-mode (see \cref{sec:CoolingTheZetaMode}).}
    \label{fig:fidelity_plots}
\end{figure}

Setting $E_L/\hbar\omega_p=10^{-3}$, we compute the gate infidelity as a function of $E_J$ and $E_{C_{\theta}}=e^2/2C_{\theta}$, see \cref{fig:fidelity_plots} (a). Note that the chosen range of parameters and the value of $E_L$ corresponds to a moderate-to-deep \zp{} regime. With these choices, we find gate fidelities between $99.99\%$ and $99.9\%$ for a broad range of system parameters, with decreasing values for increasing $E_J$ and $E_{C_{\theta}}$. This effect can be understood by contrasting the results of \cref{fig:fidelity_plots} (a) with the qubit energy level structure. In fact, we find that the gate performs better for circuit design parameters leading to increased groundstate degeneracy and moderate effective potential barriers. We give an explanation for this in \cref{sssec:Multilevel}, where we show that these conditions results in multilevel excursions limited to very few excited doublets.  

The logical action of the $\pi$ translation is shown in \cref{fig:fidelity_plots} (b), where the angle $\varphi_{\text{XZ}}$ characterizes the qubit rotation performed on the Bloch sphere in the $XZ$ plane. Note that the azimuthal angle $\varphi_{\text{XY}}$ is not shown, as it remains approximately zero with deviations smaller than $10^{-5}$. We observe that, as a function of the qubit design parameters, the gate interpolates continuously from Pauli $X$ for large $E_J/E_{C_\theta}$ to Pauli $Z$ for smaller $E_J/E_{C_\theta}$. This feature is the result of hybridization between the groundstate wave functions, as discussed above and illustrated in~\cref{fig:wave_functions_and_gate}.

\subsection{Gate fidelity with respect to circuit element disorder in $C$ and $E_L$}
\label{sssec:QuantumNOTGateFidelityDisorder}

We next study the gate behavior in presence of realistic circuit element disorder leading to coupling of the \zp{} qubit to the $\zeta$-mode. Circuit disorder also prevents independent control of the circuit degrees of freedom, and implies a parasitic drive acting on $\zeta$ when $\theta$ is driven for $dC\neq 0$. Given that the $\zeta$-mode is the main qubit decoherence channel, in this section we compute the gate fidelity including dissipation. Recall that Purcell relaxation and dephasing by photon-shot noise arise as a consequence of the parasitic coupling of $\{\phi,\theta\}$ to $\zeta$ \cite{groszkowski2017coherence}.

To treat this case, relaxation and dephasing are included into a Lindblad-form master equation, which is integrated in superoperator form for the configurations identified as A, B and C in \cref{fig:fidelity_plots} (a). The results are shown in \cref{fig:fidelity_plots} (c) for disorder in $E_L$ and in \cref{fig:fidelity_plots} (d) for disorder in $C$. In these simulations, qubit dephasing and relaxation rates are computed using the theory developed in Ref.~\cite{groszkowski2017coherence}. We consider a worst case scenario by using the maximum of the dephasing, relaxation and excitation rates obtained in full $\varphi_{\text{ext}} \in [0,2\pi]$ and $n^{\theta}_g \in [-1/2,1/2]$ excursions. The photon-loss rate $\kappa_{\zeta}$ of the $\zeta$-mode is evaluated as a function of the mode's frequency, assuming a quality factor of $Q_{\zeta}=30000$ \cite{geerlings2012improving}. Moreover, we assume a temperature of $15\,\text{mK}$. We note that taking into account the $\zeta$-mode thermal population at dilution refrigerator temperatures would lead to a photon population prohibitively large for numerical simulations. Therefore, we assume here this mode being cooled using the strategy proposed in \cref{sec:CoolingTheZetaMode}. The gate fidelity in (c) and (d) is computed with respect to the closest qubit unitary determined in \cref{sssec:QuantumNOTGateFidelityParams} in absence of circuit element disorder and dissipation. We have verified that the result does not change when the cooling power is continuously varied, and thus with the $\zeta$-mode effective thermal population up to an average of five photons.

We find that, as a consequence of a fast Hamiltonian dynamics, the gate fidelity is almost unaffected by the relatively slow dephasing and relaxation rates and that circuit element disorder is the limiting factor. We note that, despite a small to moderate degradation of the gate fidelity for disorder below $10\%$, leakage errors are appreciable for higher disorder values. Moreover, the fast gate unitary dynamics poses a control challenge. As the gate operates at a frequency which is roughly one order of magnitude smaller than the plasma frequency, the necessary time-resolution must match such a time-scale within the capability of commercially available arbitrary waveforms generators \cite{raftery2017direct,czylwikextreme}. Optimal control techniques such as GRAPE could be useful to further improve the gate fidelity, but this may require even finer time-resolution and thus be rather challenging~\cite{khaneja2005optimal,schulte2011optimal,boutin2017resonator}. 

The single-qubit gate fidelity is also found to be remarkably robust to the detailed form of the voltage pulse, moderate deviations in the external flux, and disorder in $E_J$ and $C_J$. A study of these effects is provided in~\cref{ssec:QuantumNOTGateStabilityAndExcursion}.

\subsection{Multilevel excursion during gate time}
\label{sssec:Multilevel}

\begin{figure}
    \centering{}
    \includegraphics[scale=1.0]{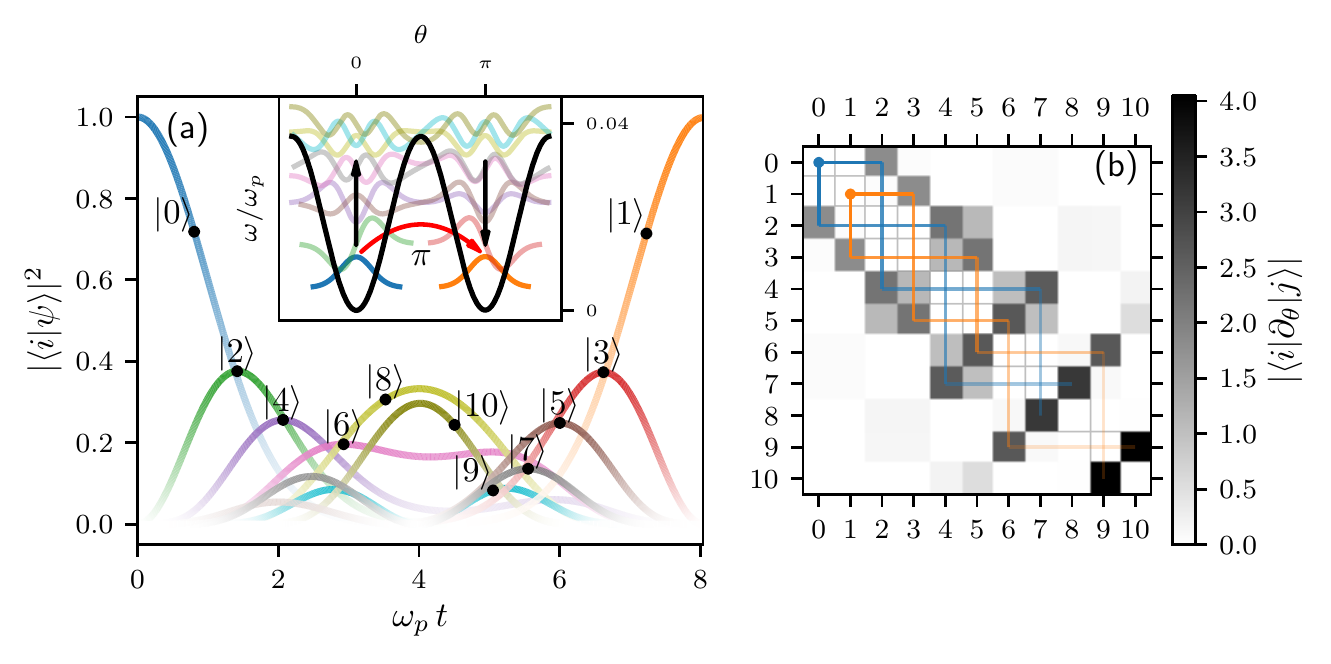}
    \caption{Multilevel excursion during the gate time. (a) State population as a function of $\omega_p t$, with initial condition $\ket{0}$. Level transparency weighted by the state population has been introduced to facilitate viewing. The insets shows the corresponding wave functions within the effective 1D model. There, black arrows illustrate how the qubit population (initially in the ground state) is transfered to higher energy doublets bridging the two potential wells, and finally transfered back to the excited state. (b) Matrix elements proportional to the charge operator $q_{\theta}$. There exist two disjoint paths connecting the ground and excited states to the higher energy doublets. We observe that the state population closely follows such paths for the few first excited levels, both at the beginning and at the end of the gate. Doublets with a higher degree of hybridization, such as (4,5) and (6,7), make it possible to transition between these two paths and from one potential well to the other. Qubit parameters $(E_L/\hbar\omega_p,E_{C_{\phi}}/\hbar\omega_p,E_{C_{\theta}}/\hbar\omega_p,E_J/\hbar\omega_p) = (10^{-3}, 0.378, 1.75\times 10^{-4},0.165).$}
    \label{fig:gate_dyson}
\end{figure}

As stated above, the proposed gate exploits the multilevel structure of the \zp{} qubit. In this section, we qualitatively discuss how this multilevel excursion takes place and the effect of leakage errors on the gate fidelity. Considering the initial state $\ket{0}$, \cref{fig:gate_dyson} (a) shows the eigenstates population as a function of time as obtained by numerical integration under the Hamiltonian \cref{eq:qubit_hamiltonian_theta_drive}. There, we observe how the initial $\ket{0}$ population is transfered by means of the voltage drive to higher energy doublets that bridge the two \zp{} potential wells. We note that the qubit population is almost completely restored to the qubit subspace at time $t=t_g$, leaving the qubit in the state $\ket{1}$.

The multilevel excursion in \cref{fig:gate_dyson} (a) can be partially anticipated by considering the matrix elements of the \zp{} charge operator $q_{\theta}$, as shown in \cref{fig:gate_dyson} (b). There, we observe a clear path leaving the ground state through levels 2 and 4, and arriving to the excited state through levels 3 and 5 after going through higher excited states. During the gate time, levels which are part of doublets with higher wave function hybridization make it possible to transfer the population between the two potential wells. Numerical experiments have shown that the number of doublets involved in the transition from $\ket{0}$ to $\ket{1}$ gives a qualitative estimate of the gate fidelity: because it leads to reduced leakage, qubit design parameters leading to excursions involving fewer levels exhibit larger fidelities. Since the number of occupied doublets grows with the height of the double-well energy barrier ($\propto E_J$), longer multilevel excursions also explain the decrease in gate fidelity observed in \cref{fig:fidelity_plots} (b). 

\subsection{Tuning the gate from $X$ to $Z$ for greater qubit control}

The continuity of the gate rotation angle as a function of the system parameters could be used to obtain a larger set of single-qubit gates. In principle, this could be achieved by adiabatically sweeping $E_{J}$ (\emph{e.g.}, replacing single junctions by tunable squid loops) or varying $\varphi_\text{ext}$ from zero to $\pi$. However, the adiabatic condition is difficult to satisfy in the qubit subspace,  requiring sweep times as large as a few milliseconds for a device in the deep \zp{} regime. Pulse shaping and optimal control techniques~\cite{berry2009transitionless,motzoi2013improving} might offer an alternative to adiabatic sweeps and need to be explored further.

\section{Fighting photon shot noise by cooling the $\zeta$-mode}
\label{sec:CoolingTheZetaMode}

For realistic circuit parameters in near-term experiments, a limiting factor for the qubit coherence times, and thus also gate and readout fidelities, is spurious coupling to the low-frequency $\zeta$-mode~\cite{groszkowski2017coherence}.
We now discuss a method to enhance the coherence times of the \zp{} qubit by cooling this mode.

In Ref.~\cite{groszkowski2017coherence}, we have shown that thermal photon population in the low-frequency $\zeta$-mode limits the coherence time of \zp{} qubits with realistic circuit parameters. To reduce the impact of this type of noise, it is essential to minimize the circuit element disorder leading to parasitic coupling of the qubit degrees of freedom to the $\zeta$-mode. Moreover, if a device can be built in the deep \zp{} regime, we have shown that there exists a threshold value $Z_\phi > Z_\text{threshold}$, such that the qubit will be protected from photon shot noise, even with circuit disorder \cite{groszkowski2017coherence}. Here, however, we consider an active approach to mitigate this problem. We engineer a protocol to boost the coherence time by cooling the $\zeta$-mode using an additional frequency-tunable resonator. Importantly, this scheme should be applicable to near-term, more realistic parameter regimes. 

\subsection{\zp{} qubit dephasing time with a cooled $\zeta$-mode}
\label{ssec:CoolingTheZetaModeGeneralTheory}

Cooling of an oscillator by periodically modulating its linear coupling to a second, heavily damped mode has been studied in the context of nanomechanical resonators \cite{tian2009ground}. There, the periodical modulation of the coupling leads to sideband transitions between the two modes, allowing for excitation of the first mode to be damped by the second. This approach is not directly applicable to our system since, as discussed in \cref{sec:CouplingStrategies}, we restrict ourselves to the use capacitors as coupling elements. We therefore propose a modification of the protocol of Ref.~\cite{tian2009ground} which relies, instead, on frequency modulation of the heavily damped mode. In practice, modulating this mode frequency also leads to a modulation of the coupling strength. Below, we develop a theory accounting for both modulated quantities, and we find that efficient cooling of the $\zeta$-mode is possible with realistic circuit parameters.

We consider an additional frequency-tunable resonator capacitively coupled to the \zp{} circuit and addressing the $\zeta$-mode as 
specified in \cref{table:coulping_the_0-pi_to_a_resonator}. The Hamiltonian for the coupled oscillators is 
\begin{equation}
\begin{split}
H_{\text{cooling}} = \hbar\omega_{\zeta} a^{\dag}a + \hbar \omega_b(t) b^{\dag}b - \hbar g(t) (a^{\dag} - a)(b^{\dag}-b),
\end{split}
\label{eq:cooling_Hamiltonian}
\end{equation}
where $a$ and $b$ are, respectively, the $\zeta$- and external- mode annihilation operators. The omission of the qubit degrees of freedom $\{\phi,\theta\}$ in \cref{eq:cooling_Hamiltonian} is justified below. The time-varying coupling constant,
\begin{equation}
\begin{split}
g(t) = \frac{C_g}{C_{\zeta} C_b}\frac{1}{2\sqrt{Z_{\zeta} Z_b(t)}},
\end{split}
\label{eq:cooling_coupling_constant}
\end{equation}
takes into account the coupling capacitance $C_g$ between the two modes as well as the capacitances $C_{\zeta}$, $C_{b},$ and the impedances $Z_{\zeta}$, $Z_{b},$ of the $\zeta$- and $b$- modes. The time dependence of the resonator frequency and the coupling strength in \cref{eq:cooling_Hamiltonian,eq:cooling_coupling_constant} is assumed to arise from the flux modulation of a tunable inductance $L_b[\Phi(t)]$ forming the $b$-mode. In particular, we assume $\omega_b(t)=\bar{\omega}_{b} + \varepsilon \cos(\omega_m t),$ where $\bar{\omega}_{b}$, $\varepsilon$ and $\omega_m$ are, respectively, the mean value, modulation amplitude and modulation frequency of the $b$-mode frequency. Accordingly, the time dependence of the coupling strength takes the form $g(t)=\bar{g}\qty[1 + \frac{\varepsilon}{2\bar{\omega}_b} \cos(\omega_m t)],$ up to first order in deviations of $L_b$ from its mean value. 

We derive an effective master equation for the $\zeta$-mode by imposing the constraint $\bar{g}\ll\omega_{\zeta},\;\omega_b$, which allows to treat the two modes as independently coupled to their respective baths. The mean frequency of the $b$-mode is chosen such that thermal excitation can safely be ignored ($\hbar\bar{\omega}_b\gg k_B T$). Moreover, the strength of the coupling between the $b$-mode and its reservoir is assumed to be frequency-independent in the range covered by the frequency modulation. Under these assumptions, the master equation of the system reads
\begin{equation}
\dot{\rho} = -\frac{i}{\hbar}[H_{\text{cooling}},\rho] + \kappa_{\zeta}[n_{\text{th}}(\omega_{\zeta})+1]\mathcal{D}[a]\rho + \kappa_{\zeta}n_{\text{th}}(\omega_{\zeta})\mathcal{D}[a^{\dag}]\rho + \kappa_{b}\mathcal{D}[b]\rho,
\label{eq:cooling_master_eq_lab_frame}
\end{equation}
where $\kappa_{\zeta}$ and $\kappa_b$ are the respective photon loss rates of the $\zeta$- and the $b$- mode, while $n_{\text{th}}(\omega_{\zeta})=1/(e^{\hbar\omega_{\zeta}/k_B T}-1)$ is the thermal photon number of the $\zeta$-mode.
$\mathcal D[x]\rho = x\rho x\dg - \frac{1}{2} x\dg x \rho - \frac{1}{2}\rho x\dg x$ is the usual dissipative superoperator.
To activate sideband transitions between the two systems, we choose the modulation frequency to be $\omega_m = \bar{\omega}_b-\omega_{\zeta}$. This choice allows for the up-conversion mechanism where photons, initially populating the $\zeta$-mode, are transfered to the external resonator and then lost to the environment at a rate $\kappa_b$. Because the external mode remains approximately in the vacuum state at all times, the inverse process is highly suppressed \cite{tian2009ground}. 

Assuming the $b$-mode to be low-Q, we employ the technique of adiabatic elimination to remove this mode from the above master equation. As discussed in more details in \cref{app:CoolingDetails}, this leads to the reduced master equation
\begin{equation}
{\dot{\rho}}^{\zeta}_I(t) = \qty[\kappa_{\zeta} (n_{\text{th}}(\omega_{\zeta}) + 1) + \Gamma_{\downarrow}]\mathcal{D}[a]{\rho}^{\zeta}_I(t) + \qty[\kappa_{\zeta} n_{\text{th}}(\omega_{\zeta}) + \Gamma_{\uparrow}]\mathcal{D}[a^{\dag}]{\rho}^{\zeta}_I(t),
\label{eq:zeta_master_eq_interaction_frame}
\end{equation}
in the interaction frame defined from~\cref{eq:cooling_Hamiltonian}. In this expression, we have defined the effective rates $\Gamma_{\downarrow} = \frac{4g'^{2}}{\kappa_b}$ and $\Gamma_{\uparrow} = \frac{4g'^{2}}{\kappa_b}\Big/{\qty[\qty(\frac{2\omega_{\zeta}}{\kappa_b/2})^2+1]}$, expressed in terms of the effective coupling strength
\begin{equation}
g' = \bar{g}\qty{J_1\qty(\frac{\varepsilon}{\omega_m})-\frac{\varepsilon}{4\bar{\omega}_b}\qty[J_0\qty(\frac{\varepsilon}{\omega_m})+J_2\qty(\frac{\varepsilon}{\omega_m})]},
\label{eq:cooling_coupling_constant_rot_frame}
\end{equation}
where $J_k(x)$ is a Bessel function of the first kind. In accordance with our assumptions, the validity of \cref{eq:zeta_master_eq_interaction_frame} is subject to the condition $\kappa_b \gg g'$. Assuming the $\zeta$-mode to be in a thermal state, the steady-state photon population under \cref{eq:zeta_master_eq_interaction_frame} is given by
\begin{equation}
\bar{n}_{\zeta}^{\text{s}} = \frac{\kappa_{\zeta}}{\gamma_{\text{cooling}}} n_{\text{th}}(\omega_{\zeta}) + \frac{\Gamma_{\uparrow}}{\gamma_{\text{cooling}}},
\label{eq:photon_number_ss}
\end{equation}
where
\begin{equation}
\gamma_{\text{cooling}} = \kappa_{\zeta} + \Gamma_{\downarrow} - \Gamma_{\uparrow}
\label{eq:cooling_rate}
\end{equation}
\noindent is the cooling rate of our scheme \cite{marquardt2008quantum}. Thermal equilibrium is therefore reached in a time $t_{\text{cooling}} = 1/\gamma_{\text{cooling}}$. 

In order to show the impact of this protocol on the coherence time of the \zp{} qubit, we follow Ref.~\cite{rigetti2012superconducting} to obtain the photon-shot noise dephasing rate for the master equation of \cref{eq:zeta_master_eq_interaction_frame}. In the limit $\chi_{01}^{\zeta}\ll\gamma_{\text{cooling}}$, this rate takes the form
\begin{equation}
\Gamma_{\varphi}^{\text{SN}} \simeq \frac{4(\chi^{\zeta}_{01})^2}{\gamma_{\text{cooling}}}\bar{n}_{\zeta}^{\text{ss}}(\bar{n}_{\zeta}^{\text{ss}}+1),
\label{eq:photon_shot_noise_rate}
\end{equation}
where we note that $\Gamma_{\varphi}^{\text{SN}}\propto \gamma_{\text{cooling}}^{-2}$ for $\bar{n}_{\zeta}^{\text{ss}}\ll 1.$ We do not find improvements on $\Gamma_{\varphi}^{\text{SN}}$ in the inverse limit $\chi_{01}^{\zeta}\gg\gamma_{\text{cooling}}$. Consequently, we observe that our cooling scheme significantly enhances the device's coherence times as long as the dispersive coupling to the $\zeta$-mode is not too large. 

\begin{figure}
    \centering
    \includegraphics[scale=1.0]{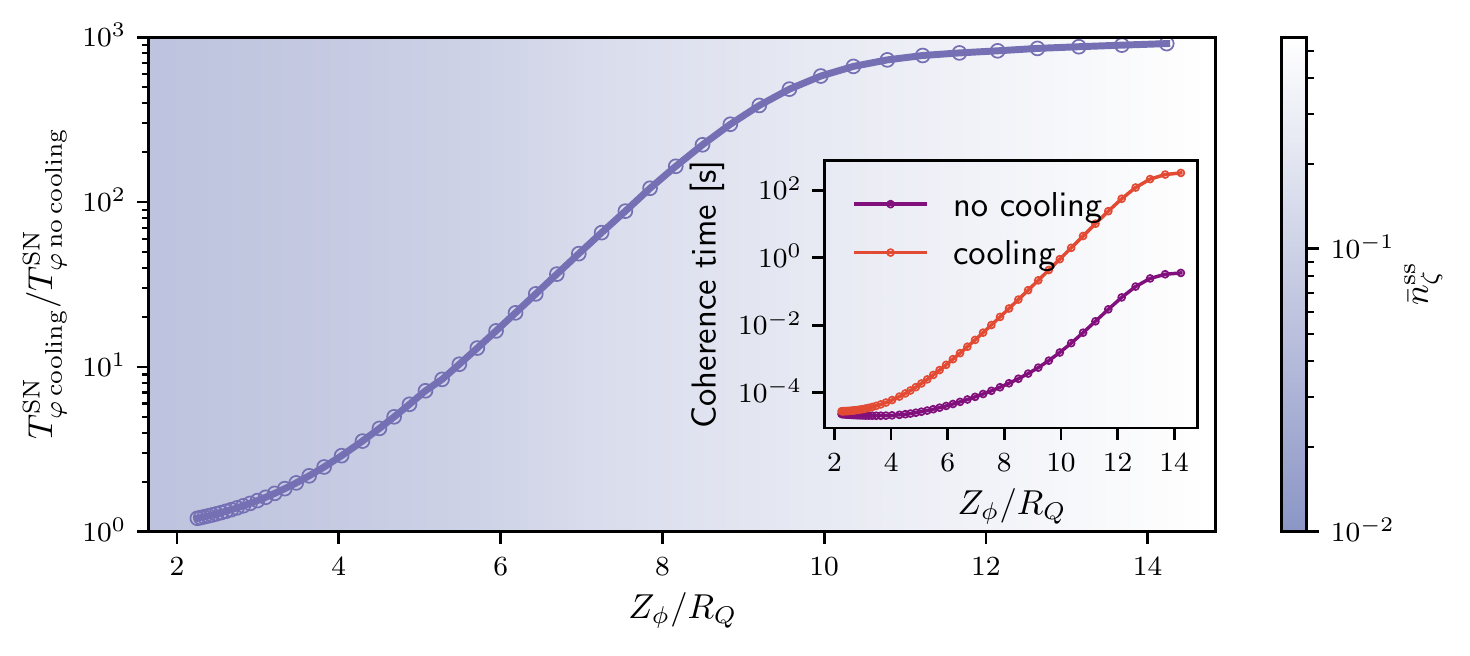}
    \caption{Photon-shot noise coherence time ($T_{\varphi}^{\text{SN}} = 1/\Gamma_{\varphi}^{\text{SN}}$) with and without cooling of the $\zeta$-mode. The inset displays the respective absolute values. Devices that can be fabricated with today's superconducting technology would be situated at the left side of this plot. Next-generation devices are expected to range between the left and the middle of the plot, where improvements on the coherence time vary between one and two orders of magnitude. In the deep \zp{} regime (to the right), major improvements will result in other noise mechanism (potentially flux noise) to be dominant. The inset displays the coherence time with and without cooling. The background density plot shows the steady-state population of the $\zeta$-mode, $\bar{n}_{\zeta}^{\text{s}}$. We note that the increase in this quantity as one moves to the deep \zp{} regime is due to the decrease of the $\zeta$-mode frequency and effective coupling to the $b$-mode, thus compromising ground state cooling. The cooling power, however, is enough to considerably reduce the dephasing rate. Circuit parameters: $E_L/\hbar\omega_p\in[1.25\times 10^{-4}, 5\times 10^{-3}]$ and $(E_{C_{\phi}}/\hbar\omega_p,E_{C_{\theta}}/\hbar\omega_p,E_J/\hbar\omega_p) = (0.25, 0.5\times 10^{-3},0.25)$, $\varepsilon/2\pi=200\,\text{MHz}$ (compatible with a SQUID array frequency-tunable resonator \cite{castellanos2007widely,stockklauser2017strong}), $\omega_b/2\pi=5\,\text{GHz}$, $Q_{\zeta}=30000$, and $T=15\,\text{mK}$.}   
    \label{fig:coolingRates}
\end{figure}

The cooling protocol is therefore applicable in a moderate-to-deep \zp{} regime, where we find improvements on the dephasing rate of the qubit by up to three orders of magnitude. The coherence time improvement due to cooling is shown in \cref{fig:coolingRates} as function of $Z_{\phi}/R_Q$. We note that the circuit parameters are the same as those in \cref{fig:exciton} (b), and correspond to the set defined as PS2 (moderate \zp{} regime) in Ref.~\cite{groszkowski2017coherence}, varying the superinductance value between those in the sets PS1 (deep \zp{} regime) and PS3 (near-term regime) of the same paper. As anticipated, the relative gain becomes significant as one moves towards the deep \zp{} regime (large $Z_{\phi}/R_Q$), before reaching saturation. The saturation value can can be understood from \cref{eq:photon_shot_noise_rate} in the limit of $n_{\text{th}}(\omega_{\zeta})\rightarrow\infty$, where it is only a function of $\bar{g}$ and $\varepsilon/\bar{\omega}_b$.

The interaction with the $b$-mode further broadens the $\zeta$-mode, resulting in larger \zp{} qubit Purcell relaxation and excitation rates. Given that such rates have been found not to limit the qubit coherence, we do not expect this effect to be a limiting factor in practice~\cite{groszkowski2017coherence}. In fact, we predict the increase of the Purcell rates to be one order of magnitude, which is still far from compromising the device. 

\subsection{Effect of parasitic coupling and implementation details}
\label{ssec:Implementation}

Circuit element disorder responsible for the coupling between the qubit degrees of freedom and the $\zeta$-mode also introduces a parasitic coupling between $\{\phi,\theta\}$ and the $b$-mode. As a result, the fact that $\omega_m$ is specially chosen to activate a resonant interaction between the $\zeta$-mode and the frequency-tunable device, implies that any qubit transition matching $\omega_{\zeta}$ will also be resonant. Given that, by design, the \zp{} qubit transition should not be resonant with the $\zeta$-mode, accidental resonances might arise within the multilevel structure of the device. This possibility, however, can be minimized by circuit design. Additionally, resonances between the \zp{} circuit transitions and the mean frequency of the $b$-mode should be avoided with by properly choosing $\bar{\omega}_b$.

Finally, we discuss some of the implementation details leading to a correction of the modulation frequency. We first address the effect the dispersive interaction between the qubit degrees of freedom and the $\zeta$-mode~\cite{Dempster2014a,groszkowski2017coherence}. In the limit $\chi_{01}^{\zeta}\ll\gamma_{\text{cooling}},$ the frequency of the $\zeta$-mode is approximately independent of the qubit state. Therefore, we account for the mean $\zeta$-mode frequency shift due to the dispersive interaction by redefining the $b$-mode modulation frequency as
\begin{equation}
\omega_m \to \omega_m - (\chi_0^{\zeta}+\chi_1^{\zeta})/2,
\label{eq:effective_modulation_frequency}
\end{equation} 
where $\chi^{\zeta}_{0}$ ($\chi^{\zeta}_1$) is the dispersive shift for the qubit being in the ground (excited) state \cite{groszkowski2017coherence} [see also \cref{eq:Hamiltonian_Dispersive}]. A similar effect is expected to arise from nonlinear terms in the $b$-mode Hamiltonian. In fact, it is worthwhile to note that current implementations of frequency-tunable resonators rely on Josephson junctions which introduce a small Kerr nonlinearity, $K$, and comparable shift to $\omega_b$ \cite{sandberg2008tuning,hatridge2011dispersive,eichler2014quantum,castellanos2007widely,stockklauser2017strong}. Given that, by design, the $b$-mode is kept in a nearly vacuum state at all times during the cooling protocol, the effect of the nonlinearity is limited to a frequency shift. The latter can again be compensated by changing the modulation frequency according to $\omega_m\to\omega_m-K/2$. We note that the results of this section, including the reduced master equation \cref{eq:zeta_master_eq_interaction_frame} and the effect of nonlinearities, were validated against the integration of the full time-dependent master equation of \cref{eq:cooling_master_eq_lab_frame}.

\section{Conclusion}
\label{sec:Conclusions}

The \zp{} circuit is a promising candidate for the realization of a protected superconducting qubit. However, both fabrication and control challenges need to be overcome. In this paper, we considered control strategies exploiting the multilevel structure of this device, within a realistic circuit model.

We explored the possibility of dispersively coupling the \zp{} qubit to a resonator, which can be used for standard dispersive readout and resonator mediated gates. In general, dispersive coupling is extremely small in the moderate-to-deep \zp{} regime due to the highly symmetric double-well structure of the qubit potential. Nevertheless, we found a remarkably large dispersive shift by coupling to the $\phi$ mode of the \zp{} qubit, and operating in a regime reminiscent of the straddling regime of a transmon qubit. Dispersive shifts around a hundred kHz could be achievable, even rather deep in the \zp{} regime. This is promising for qubit characterization through spectroscopy, and might also be promising for readout and gates due to the extremely long qubit lifetimes that are possible in this regime.

We moreover proposed a new, fast and high-fidelity single-qubit gate that can smoothly interpolate between logical $X$ and $Z$ by varying the qubit operation point. We studied the gate fidelity as a function of the \zp{} circuit and control parameters, and the amount of circuit element disorder. We found that the gate fidelity is not significantly affected for small deviations from optimal parameters and moderate disorder. Future work will concentrate on extending the gate operation to a universal set of single-qubit gates. Finally, we note that qubits with a similar level structure to that of the \zp{} qubit might leverage similar ideas~\cite{douccot2012physical,Kitaev01,Lutchyn10,Oreg10,huang2015manipulating,Aasen16,petrescu2017fluxon}. 

In addition, we have designed a protocol to enhance the qubit coherence time due to photon shot noise from the $\zeta$-mode. Our scheme couples this mode to a frequency-modulated and highly damped resonator, which is used as a zero temperature-bath for the $\zeta$-mode. We characterized the improvement in the qubit photon shot noise dephasing time as a function of the circuit design parameters. While the coherence time enhancement for near-term devices is expected to provide a 2 to 10 times gain on coherence, we predict improvements of one and two orders of magnitude for the future generations. We also envision that this active cooling protocol could be useful in a more general context of superconducting devices with low-frequency modes or with residual thermal population. 

Several open questions remain about how to best use the ingredients presented in this paper for a \emph{universal} set of logical operations. A quantitative analysis is needed to determine the potential use of dispersive coupling for readout and gates, taking into account any possible degradation of qubit coherence due to coupling to the resonator, and whether gate times can be sufficiently fast compared to the coherence times to achieve high fidelity gates. It would also be interesting to exploit the tunability of the gate introduced in~\cref{sec:QuantumNOTGate} to achieve a larger set of single qubit gates. Finally, the original proposal for protected phase gates from Ref.~\cite{Brooks2013} should be investigated in a more realistic setting, to determine the potential use of this approach for near-to-medium term experiments.

\section{Acknowledgments}
\label{sec:Acknowledgments}

We acknowledge valuable discussions with Andr\'as Gyenis, Samuel Boutin and Christian K. Andersen. ADP acknowledges support from the Fundaci\'on Williams en Argentina and the Bourse d'excellence de 3e cycle, Facult\'e des Sciences, Universit\'e de Sherbrooke. This work was supported by the Army Research Office under Grant no. W911NF-15-1-0421 and NSERC. This research was undertaken thanks in part to funding from the Canada First Research Excellence Fund. This work is supported by the Australian Research Council (ARC) via Centre of Excellence in Engineered Quantum Systems (EQUS) Project No. CE170100009.

\appendix
\section{Circuit Hamiltonian in presence of gate and ground capacitance disorder}
\label{app:gate_and_ground_assymetries}

The effective capacitances $C_\mu$ of the \zp{} circuit modes $(\phi,\theta,\zeta,\Sigma)$ introduced in~\cref{eq:0pi_hamiltonian} are given by
\begin{equation}
\begin{split}
&C_{\phi} = {C_0+C_g} + 2C_J,\\
&C_{\theta} = {C_0+C_g} + 2(C + C_J),\\
&C_{\zeta} = {C_0+C_g} + 2C, \\
&C_{\Sigma} = {C_0+C_g}.\\
\end{split}
\label{eq:mode_capacitances}
\end{equation}
Using definitions found in \cref{sssec:VoltageDriveCoupling}, the full expression of the term $H_{dC_g,dC_0}$ in \cref{eq:Hcoupl} is
\begin{equation}
\begin{split}
H_{dC_g,dC_0}=&-\sum_{\mu}\frac{1}{2C_{\mu}^2}\qty[(C_g/2)dC_{g_{\Sigma}}+(C_0/2)dC_{0_{\Sigma}}]q_{\mu}^2\\
&-\frac{1}{2}\sum_{\mu\neq\nu\neq\sigma \text{ all }\neq \Sigma}\frac{1}{C_{\mu}C_{\nu}}\qty[(C_g/2)dC_{g_{\sigma}}+(C_0/2)dC_{0_{\sigma}}]q_{\mu}q_{\nu}\\
&-\sum_{\mu\neq\Sigma} \frac{1}{C_{\Sigma}C_{\mu}}\qty[(C_g/2)dC_{g_{\mu}}+(C_0/2)dC_{0_{\mu}}]q_{\mu}q_{\Sigma},
\end{split}
\label{eq:dCg_dC0_kinetic_energy}
\end{equation}
Disorder of the from $dC_{g_{\mu}},\,dC_{0_{\mu}}\neq 0,$ is assumed to be small compared to all other capacitances in the circuit, and $H_{dC_g,dC_0}$ is therefore neglected in this work.
The expression for $H_\text{drive}^\text{asymm}$ in~\cref{eq:qubit_hamiltonian} is
\begin{equation}
\begin{aligned}
H_\text{drive}^\text{asymm} ={}&
-\frac{C_g\,C_J\,dC_J}{C_{\phi}C_{\theta}}\qty(V_{\theta}q_{\phi}+V_{\phi}q_{\theta})-\frac{C_g\,C\,dC}{C_{\zeta}C_{\theta}}\qty(V_{\theta}q_{\zeta}+V_{\zeta}q_{\theta}) \\
&+\sum_{\mu\neq\Sigma}\frac{C_g}{C_{\mu}^2}\qty[(C_{\mu}-C_g/2)dC_{g_{\Sigma}}-(C_0/2)dC_{0_{\Sigma}}]V_{\mu}q_\mu\\
&+\sum_{\mu\neq\Sigma}\frac{C_g}{C_{\mu}C_{\Sigma}}\qty[(C_{\Sigma}-C_g/2)dC_{g_{\mu}}-(C_0/2)dC_{0_{\mu}}]V_{\Sigma} q_{\mu}\\
&+\sum_{\mu\neq\nu\neq\sigma \text{ all }\neq \Sigma}\frac{C_g}{C_{\mu}C_{\nu}}\qty[(C_{\nu}-C_g/2)dC_{g_{\sigma}}-(C_0/2)dC_{0_{\sigma}}]V_{\nu}q_{\mu} \\
&+ \sum_{\mu}\frac{C_g}{C_{\mu}C_{\Sigma}}\qty[(C_{\mu}-C_g/2)dC_{g_{\mu}}-(C_0/2)dC_{0_{\mu}}]V_{\mu}q_{\Sigma}.
\end{aligned}
\end{equation}
In these two expressions, $V_{\mu},$ $dC_{g_{\mu}}$ and $dC_{0_{\mu}}$ are given in terms of $V_{i},$ $dC_{g_{i}}$ and $dC_{0_{i}}$, respectively, according to the transformation rule specified in~\cref{eq:normal_modes} (where Greek indices denote the normal mode variables and Latin indices node variables).

\section{One-dimensional effective model}
\label{app:1d_effective_Hamiltonian}

The reduction from the \zp{} Hamiltonian to a 1D effective model was first motivated in Ref.~\cite{Brooks2013} and analytically studied in Ref.~\cite{shen2015theoretical} in the context of the Born-Oppenheimer approximation. In \cref{sssec:QubitStructure}, moreover, we have provided an intuitive justification for such a model. Here, we perform a numerical calculation which, in contrast to analytical approaches, does not require additional approximations. 

Starting with the \zp{} circuit Hamiltonian in absence of disorder, i.e. \cref{eq:symm_qubit_Hamiltonian}, we define
\begin{equation}
H_{\phi}=\frac{q_{\phi}^2}{2C_{\phi}} -2E_J\cos\tilde\theta\cos\qty(\phi-\varphi_{\text{ext}}/2) + E_L \phi^2,
\label{eq:phi_Hamiltonian}
\end{equation}
where $\tilde\theta \in [-\pi/2,3\pi/2)$ acts here as a parameter. This corresponds to the first step of the Born-Oppenheimer approximation, where only the less massive degrees of freedom ($\phi$ in our case) are considered. We then find the groundstate energy $E_{0}(\tilde\theta)$ of $H_{\phi}$, as a function of $\theta$. As a next step, we define a second 1D problem by the Hamiltonian
\begin{equation}
H_{\theta}=\frac{q_{\theta}^2}{2C_{\theta}} + E_0(\theta).
\label{eq:theta_Hamiltonian}
\end{equation}
where $E_{0}(\tilde\theta\to\theta)$ is used as an effective potential and $\theta$ is qubit phase operator. Note that, in contrast to \cref{eq:phi_Hamiltonian}, $H_{\theta}$ governs the motion of the massive degrees of freedom ($\theta$ in our case). 

\begin{figure}
    \centering{}
    \includegraphics[scale=1.0]{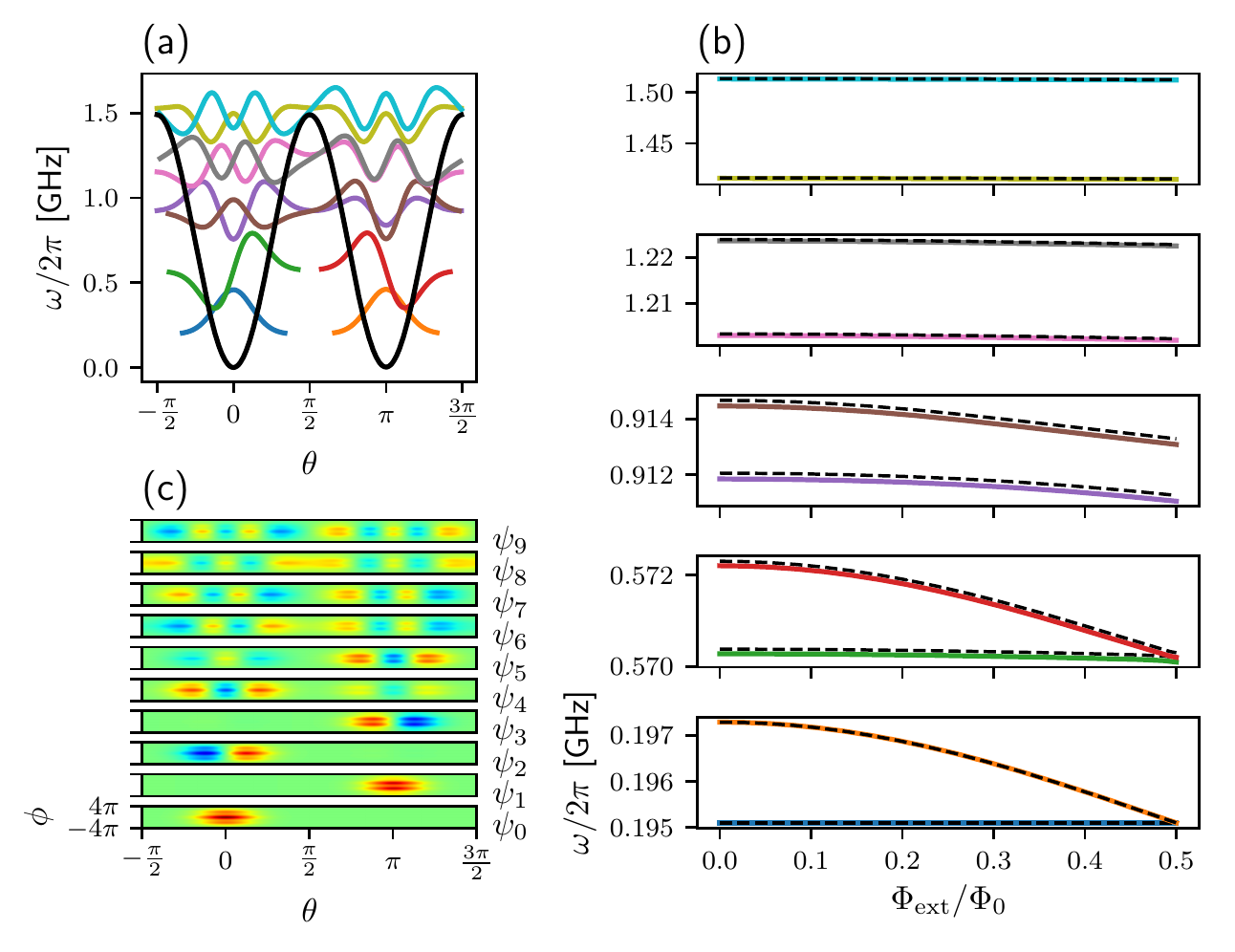}
    \caption{1D effective model for the \zp{} qubit. (a) 1D effective potential (black) and 1D wave functions (color) offset by their respective energy. The wave functions corresponding to the $H_{\theta}$ eigenstates are displayed in panel (a) for $\varphi_{\text{ext}}=0$, along with the resulting 1D effective potential. Panel (b) shows the respective eigenvalues (solid colored lines), along with the spectrum of the full \zp{} Hamiltonian \cref{eq:symm_qubit_Hamiltonian} (black dashed lines). In (c), we show the two-dimensional eigenfunctions corresponding to \cref{eq:symm_qubit_Hamiltonian} (for $\varphi_{\text{ext}}=0$), which should be contrasted with those of the 1D model in (a). The effective 1D potential has an exact $2\pi$- and approximately $\pi$-periodic structure. Qubit parameters $(E_L/\hbar\omega_p,E_{C_{\phi}}/\hbar\omega_p,E_{C_{\theta}}/\hbar\omega_p,E_J/\hbar\omega_p) = (10^{-3}, 0.378, 1.75\times 10^{-4},0.165)$ and $\omega_p/2\pi=40\,\text{GHz}$. }
    \label{fig:1D_model}
\end{figure}

Equation \ref{eq:theta_Hamiltonian} represents the one-dimensional effective Hamiltonian for the \zp{} qubit. The corresponding eigenvalues and eigenstates are shown in \cref{fig:1D_model} as a function of the external flux. Remarkably, we find excellent agreement with the complete two-dimensional circuit Hamiltonian \cref{eq:symm_qubit_Hamiltonian} for several excited doublets. By fitting the effective potential, we find that \cref{eq:theta_Hamiltonian} can very accurately be written as 
\begin{equation}
H^{\text{eff}} = 4 E_{C_{\theta}}(n_{\theta}-n^{\theta}_g)^2 - E_{2}(\varphi_{\text{ext}})\cos2\theta - E_{1}(\varphi_{\text{ext}})\cos\theta,
\label{eq:effective_1D_Hamiltonian}
\end{equation}
where we incorporate the offset charge $n_g^{\theta}$. In the above expression, the potential energy coefficients read $E_{2}(\varphi_{\text{ext}})=E_{\alpha} - E_{\beta}\cos(\varphi_{\text{ext}})$ and $E_{1}(\varphi_{\text{ext}})=E_{\gamma}\cos(\varphi_{\text{ext}}/2)$ regardless of the qubit design parameters. The relations $E_{\alpha} \gg E_{\gamma}$ and $E_{\alpha} \gg E_{C_{\theta}}$, satisfied in the deep \zp{} limit, ensure exponential suppression of relaxation and dephasing rates \cite{PhysRevLett.112.167001}. For the set of parameters in \cref{fig:1D_model}, we find $E_{\alpha}/\hbar\omega_p=1.8608\times 10^{-2},$ $E_{\beta}/\hbar\omega_p=1.0073\times 10^{-8},$ $E_{\gamma}/\hbar\omega_p=2.6625\times 10^{-5}$. Finally, we note that an expression similar to \cref{eq:effective_1D_Hamiltonian} has been theoretically proposed in Ref.~\cite{shen2015theoretical}. We have, however, found necessary to incorporate additional flux-dependence to the potential energy coefficients. 

\section{Raman processes for qubit control}
\label{app:RamanGates}

We now consider enabling a Raman-type gate operation by virtually populating the excited states of the qubit. In particular, we study the effective dynamics in the ground state manifold $\{\ket{0},\ket{1}\}$ by performing adiabatic elimination of the first few qubit excited level. As only virtual transitions to high-energy levels are involved, Raman-type gates could, in principle, preserve the device's noise protection to some degree. However, as shown below, the transition amplitude $\ket{0}\leftrightarrow\ket{1}$ vanishes in a large parameter range because of an approximate selection rule. To arrive at this result, we make use of the adiabatic elimination procedure developed in Ref.~\cite{reiter2012effective}, which applies to weakly and off-resonantly driven multilevel systems.

\subsection{Single-tone driving}
\label{sapp:SingleFreqDriv}

We consider the \zp{} circuit capacitively coupled to microwave voltage sources addressing $\theta,$ as described in the second row of \cref{table:coulping_the_0-pi_to_a_resonator}. Including a total of $M$ qubit levels, the driven \zp{} qubit Hamiltonian can be written as

\begin{equation}
H/\hbar = \sum_{i=0}^M \omega_i \sigma_{ii}+ \sum_{i=0,1}\sum_{j=2}^M (\Omega_{ij} e^{-i\omega t} \sigma_{ji} + h.c.).
\label{eq:single_drive_Hamiltonian}
\end{equation} 
Here, $\omega_i$ is the frequency of the $i^{\text{th}}$ eigenstate, $\omega$ the frequency of the drive and $\Omega_{ij}= \frac{C_g}{C_{\theta}} (eV) e^{-i\beta}\bra{j}n_{\theta}\ket{i}$ is the coupling strength between levels $(i,j)$ for a voltage pulse of the from $V_{\theta}(t)=V\cos(\omega t + \beta)$. We note that rapidly rotating terms have been dropped in \cref{eq:single_drive_Hamiltonian} under the assumption of a weak drive ($\Omega_{ij}/\omega \ll 1$). Moreover, we neglect the effect of $\Omega_{01}$, which is exponentially small for typical qubit design parameters. 

By numerically solving the Schr\"odinger equation, we find that \cref{eq:single_drive_Hamiltonian} hardly generates qubit population inversion. In fact, we observe that states from the few excited doublets destructively interfere with each other, thus leading to a negligible transition amplitude. This cancellation is preserved in a broad range of qubit design parameters, including flux excursions from the standard operating point $\varphi_{\text{ext}}=0.$  

\begin{figure}
    \centering{}
    \includegraphics[scale=1.0]{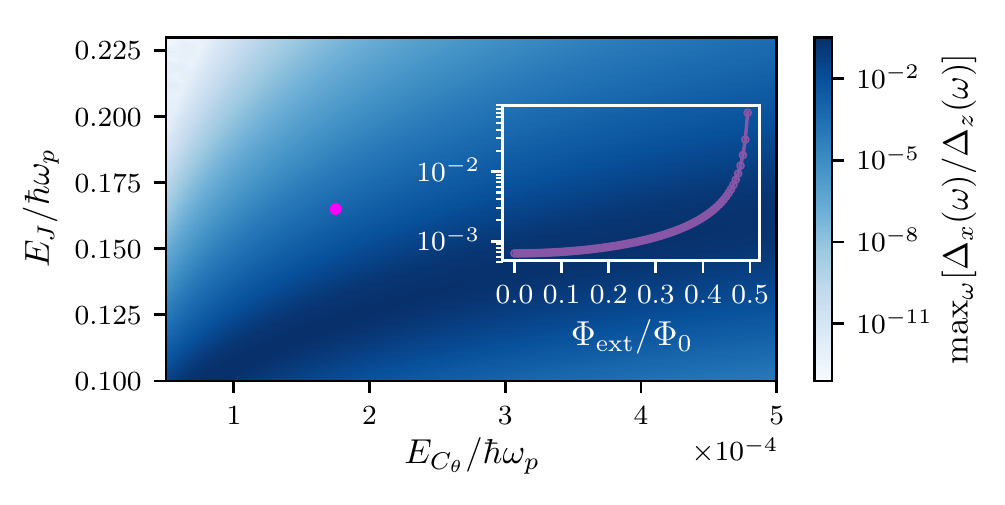}
    \caption{Ratio $\Delta_x(\omega)/\Delta_z(\omega)$ optimized over the drive frequency $\omega$ under the off-resonant drive condition, as a function of the \zp{} circuit design parameters. Inset: optimal $\max_{\omega}[\Delta_x(\omega)/\Delta_z(\omega)]$ as a function of the external flux for the parameters $(E_L/\hbar\omega_p,E_{C_{\phi}}/\hbar\omega_p,E_{C_{\theta}}/\hbar\omega_p,E_J/\hbar\omega_p) = (10^{-3}, 0.378, 1.75\times 10^{-4},0.165).$. The simulations included $30$ qubit levels.}   
    \label{fig:raman_amplitude}
\end{figure}

To understand this effect, we reduce the multilevel dynamics to the qubit subspace. Following Ref.~\cite{reiter2012effective} and modeling dissipation by the set of collapse operators $\{L_{j0} = \sqrt{\gamma_{j0}} \sigma_{0j}, L_{j1} = \sqrt{\gamma_{j1}} \sigma_{1j};\; j=2,...,M\}$, we find 
\begin{equation}
\begin{split}
H^{\text{eff}}/\hbar =& \sum_{i, i'=0,1}\qty(\omega_i \delta_{ii'} - \frac{1}{2}\sum_{j=2}^M \frac{(\Omega_{ij})^*\Omega_{i'j}\qty[\Delta_{ji}(\omega) + \Delta_{ji'}(\omega)]}{\qty[i\frac{\gamma_j}{2}+\Delta_{ji}(\omega)]\qty[-i\frac{\gamma_j}{2} + \Delta_{ji'}(\omega)]})\sigma_{ii'},
\label{eq:effective_hamiltonian_single_drive}
\end{split}
\end{equation}
where $\gamma_j = \gamma_{j0} + \gamma_{j1}$ is the total decay rate from the $j^{\text{th}}$ level to the ground state manifold and $\Delta_{ji}(\omega) = (\omega_j - \omega_i) - \omega$ is the detuning of the drive with respect to the transition frequency $\omega_j - \omega_i$. The validity of this effective model is subject to the condition of off-resonant driving ($|\Omega_{ij}/\Delta_{ji}(\omega)|\ll 1$). By taking into account the phase of the drive, the effective Hamiltonian can be written as
\begin{equation}
H^{\text{eff}}/\hbar = \frac{\Delta_z(\omega)}{2} \sigma_z + \frac{\Delta_x(\omega)}{2} \sigma_x,
\label{eq:effective_Hamiltonian_sigma}
\end{equation}
from which we expect to see Rabi oscillations for $\Delta_x(\omega)/\Delta_z(\omega) \gg 1$ (Raman gate). \cref{fig:raman_amplitude} shows this ratio as a function of the \zp{} circuit design parameters and optimized with respect to the drive frequency $\omega$ under the off-resonant drive condition. We observe that $\Delta_x(\omega)/\Delta_z(\omega)$ remains small throughout the range of parameters. Furthermore, this ratio is smaller than $10^{-4}$ when considering the \zp{} parameters of \cref{fig:raman_amplitude} with vanishing overlap between logical wave functions. Additionally, we show the result of the optimization for a full flux excursion (figure inset). Despite an appreciable increase of $\Delta_x(\omega)/\Delta_z(\omega)$ with $\Phi_{\text{ext}}/\Phi_0,$ we find this improvement not enough to allow for qubit control. 

Considering instead a drive addressing $\phi$, we perform the optimization of $\Delta_x(\omega)/\Delta_z(\omega)$ to again find, in the majority of cases, only a negligible $\sigma_x$ component in the effective Hamiltonian \cref{eq:effective_Hamiltonian_sigma}. Interestingly, we have also identified some qubit design parameters for which $H^{\text{eff}}\propto \sigma_x$. However, such configurations are sparsely distributed over the range numerically explored, and the results are sensitive to small parameter deviations. Therefore, driving $\phi$ does not appear to be a practical solution. 

\subsection{Two-tone driving}
\label{sapp:DoubleFreqDriv}

We now investigate if the presence of multiple drives could help to overcome the off-diagonal component cancellation found above, in particular by optimizing on the relative phase of these drives. In the presence of two microwave voltage pulses (labeled as $d_1$ and $d_2$), the qubit Hamiltonian reads
\begin{equation}
H/\hbar = \sum_{i=0}^M \omega_i \sigma_{ii}+ \sum_{i=0,1}\sum_{j=2}^M \sum_{k=1,2}(\Omega^{k}_{ij} e^{-i\omega^{d_k} t} \sigma_{ji} + h.c.).
\label{eq:double_drive_Hamiltonian_timeindep}
\end{equation} 
Following again the procedure in Ref.~\cite{reiter2012effective}, we find
\begin{equation}
\begin{split}
H^{\text{eff}}/\hbar =& \sum_{i, i'=0,1}\qty(\omega_i \delta_{ii'} - \frac{1}{2}\sum_{k, l=1,2}\sum_{j=2}^Me^{i\omega^{d_{kl}}t} \frac{(\Omega^{d_k}_{ij})^*\Omega^l_{i'j}\qty[\Delta_{ji}(\omega^{d_k}) + \Delta_{ji'}(\omega^{d_l})]}{\qty[i\frac{\gamma_j}{2}+\Delta_{ji}(\omega^{d_k})]\qty[-i\frac{\gamma_j}{2} + \Delta_{ji'}(\omega^{d_l})]})\sigma_{ii'},
\label{eq:effective_hamiltonian_double_drive}
\end{split}
\end{equation}
where $\omega^{d_{kl}} = \omega^{d_k} - \omega^{d_l}$ is the difference between two drive frequencies. We note that, if the two drives have the same frequency, their relative phase is factored out of the transition amplitude, thus reducing \cref{eq:effective_hamiltonian_double_drive} to \cref{eq:effective_hamiltonian_single_drive}. The most general situation, however, was investigated by numerical simulation of \cref{eq:double_drive_Hamiltonian_timeindep} and \cref{eq:effective_hamiltonian_double_drive}. None of the explored qubit design parameters and drive frequencies have shown a significant change with respect to what was found in the single-drive case.

\section{Gate fidelity as a function of control parameters and circuit element disorder in $E_J$ and $C_J$}
\label{ssec:QuantumNOTGateStabilityAndExcursion}

In this section, we study the gate fidelity accounting for the effect of pulse shaping and moderate deviations in the external flux that sets the qubit operating point. Moreover, we investigate the gate fidelity in the presence of circuit element disorder introducing additional $\phi\leftrightarrow\theta$ coupling and a spurious drive on $\phi$, without involving the $\zeta$-mode.

\subsubsection{Gate fidelity as a function of drive strength and duration}
\label{sssec:QuantumNOTGateStabilityVotageAndTiming} 

To understand how pulse shaping affects the gate performance, we first analyze the effect of the drive strength and duration. In \cref{fig:gate_behaviour} (a), we show the gate infidelity considering a square voltage pulse of amplitude $V_{\text{sq}}$ and duration $t_g.$ There, the diagonal features are high-fidelity regions. While the first of these (starting from the left) corresponds to a $\sigma_x$ gate operation, the second corresponds to an identity operation or $\sigma_x^2$. This pattern repeat itself for the subsequent pairs of features as $t_g$ increases, but with decreasing gate fidelity due to leakage errors. The nonregular spacing of the high-fidelity regions indicates that, in contrast to more standard gate schemes, the proposed single-qubit gate has a nonlinear dependence on the drive strength and the evolution time. In fact, the first high-fidelity feature has a hyperbolic shape (visible on a larger scale) defined by the relation $({C_g}/{C_{\theta}})2eV_{\text{sq}}/\hbar \times t_g\simeq \pi,$ which is derived from a short-time approximation of the gate propagator. The fidelity along such hyperbola is not constant, and there exist an optimal drive strength and gate time about which the gate fidelity is maximal and, to first order, insensitive to deviations. Such an optimal point, computed as a function of the qubit design parameters, has been used to produce the results in \cref{fig:fidelity_plots}. Away from the optimal point, however, the slow decrease of the gate fidelity along the mentioned hyperbola could be leveraged to extend the gate time if necessary for control purposes.

\subsubsection{Variations in gate fidelity for a shaped pulse}
\label{sssec:QuantumNOTGateStabilityPuslse}
We now consider the effect of a hyperbolic tangent pulse shape with a finite turn on and shutdown time $\sigma.$ In this section, the pulse area and maximum drive strength are kept constant and equal to those of the optimal square pulse. The latter is also used as a standard to obtain the relative change of the gate fidelity shown in red in \cref{fig:gate_behaviour} (b). As can be seen, the gate fidelity remains almost unchanged in a wide range of $\sigma$, including turn -on and -off times that are comparable to the total optimal square pulse length. The same conclusion applies to all the qubit design parameters considered in this work.

\subsubsection{Gate fidelity as a function of flux}
\label{sssec:QuantumNOTGateStabilityVotageFlux} 

We have so far discussed the gate fidelity for a qubit operating at $\Phi_{\text{ext}}/\Phi_0=0$ ($\varphi_{\text{ext}}= 0$). In this section, however, we investigate the effect of external flux variations. \cref{fig:gate_behaviour} (b) shows the relative change of the gate fidelity as a function of $\Phi_{\text{ext}}/\Phi_0$ (in violet). Here, the fidelity is computed with respect to a fixed gate unitary determined at $\Phi_{\text{ext}}/\Phi_0=0$, while the voltage drive parameters are kept to the optimal values determined for such a configuration. These conditions ensure that we only observe the effect of varying the external flux. We note that gate fidelity remains essentially unchanged for small to moderate flux excursions from the qubit operating point. This behavior, which is in part a consequence of the small flux dispersion of the \zp{} qubit, shows that the proposed gate will tolerate the small flux fluctuations that could arise in practice. 

Since the external flux affects the hybridization of the \zp{} logical wave functions, the former could be used to tune the gate operation similarly to what was shown in \cref{fig:fidelity_plots} (b). Choosing the qubit design parameters such that the gate implements a $\sigma_x$ operation at $\Phi_{\text{ext}}/\Phi_0=0$, a slowly varying external flux would smoothly rotate the gate operation implementing a $\sigma_z$ gate for $\Phi_{\text{ext}}/\Phi_0=0.5$ ($\varphi_{\text{ext}}=\pi$). However and as discussed above, because of the near degeneracy of the qubit states a device in the deep \zp{} regime implies adiabatic sweep times in the millisecond range, thus limiting the applicability of the gate flux tunability. 

\subsubsection{Gate fidelity in presence of $E_J$ and $C_J$ disorder}
\label{sssec:QuantumNOTGateStabilityEJandCJ} 
\begin{figure}
    \centering{}
    \includegraphics[scale=1.0]{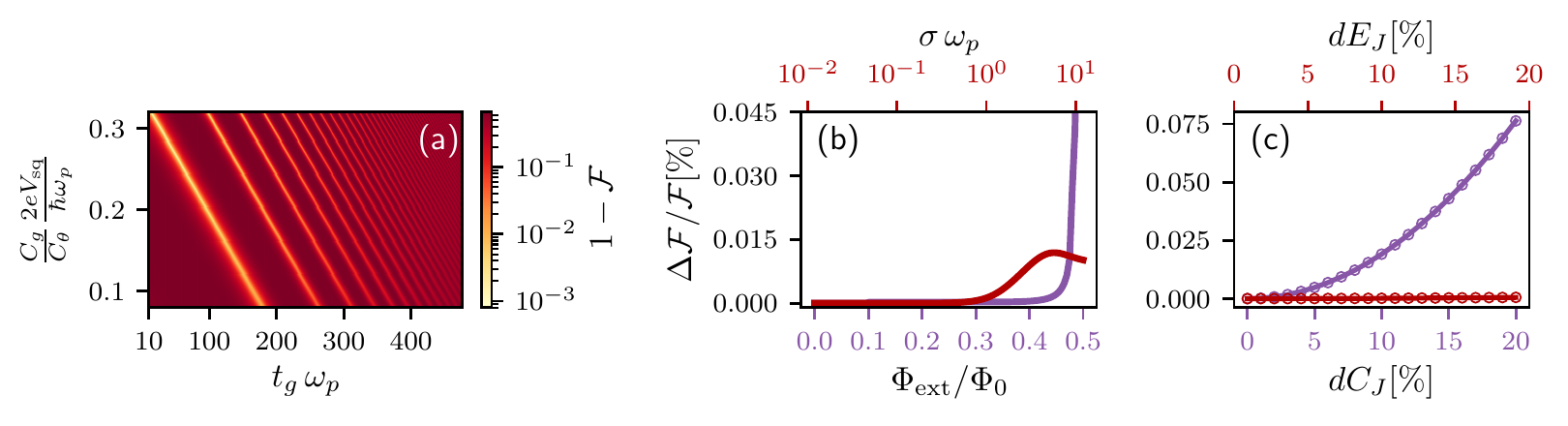}
    \caption{Variation in the gate fidelity for the \zp{} qubit design parameters $(E_L/\hbar\omega_p,E_{C_{\phi}}/\hbar\omega_p,E_{C_{\theta}}/\hbar\omega_p,E_J/\hbar\omega_p) = (10^{-3}, 0.378, 1.75\times 10^{-4},0.165).$ (a) Gate infidelity in the $(V_{\theta},t_g)$ plane. Brighter regions correspond to high-fidelity qubit operations from where the optimal drive amplitude and duration is determined (see \cref{sssec:QuantumNOTGateStabilityVotageAndTiming}). (b) Relative change in gate fidelity as a function of the voltage-drive turn -on and -off time $\sigma$ (in red), and the phase associated to the external magnetic flux $\Phi_{\text{ext}}$ (in violet). (c) Effect of disorder in $E_J$ (in red) and $C_J$ (in violet) on the gate fidelity.}   
    \label{fig:gate_behaviour}
\end{figure}

Having previously considered the effect of circuit element disorder that lead to coupling of the qubit degrees of freedom to the $\zeta$-mode, we now consider disorder leading to additional parasitic coupling between $\phi$ and $\theta$. \cref{fig:gate_behaviour} (c) shows the relative change of the gate fidelity due to disorder in $E_J$ (in red) and $C_J$ (in violet). We observe that the gate performance is only slightly affected in a realistic range of asymmetries, expected to be between $1\%$ and $10\%$. A non-zero $dC_J$ results in a parasitic drive on the $\phi$ coordinate which has also been included in the simulations. Because $E_J$ disorder does not lead to such an unwanted drive, its effect on the gate fidelity is negligible in comparison. 

\section{Master equation for a cooled $\zeta$-mode}
\label{app:CoolingDetails}

In this section, we derive the effective $\zeta$-mode master equation of \cref{eq:zeta_master_eq_interaction_frame}. This is done using an adiabatic elimination of the external resonator under the assumption $g'\ll\kappa_b$. Recall that $g'$ is the effective interaction strength between the $b$ and the $\zeta$ modes in the interaction frame. Following Ref.~\cite{marquardt2008quantum}, we treat the heavily-damped external resonator as a bath for the $\zeta$-mode. To this end, we specify the system and bath Hamiltonians, as $\hbar\omega_{\zeta} a^{\dag}a$ and $\hbar\omega_b(t)b^{\dag}b,$ respectively. Going to a frame rotating at $\omega_{\zeta}$ for the $\zeta$-mode and at the modulated frequency $\omega_b(t)$ for the $b$-mode, the  interaction Hamiltonian reads
\begin{equation}
H^{\text{int}}_I(t) = -\bar{g}\qty(a^{\dag}f^*(t) - a f(t))\qty(b^{\dag}_I(t) - b_I(t)),
\label{eq:interaction_Hamiltonian_zeta_and_resonator}
\end{equation} 
where $f(t)=e^{-i\omega_a t} + \frac{\varepsilon}{4\bar{\omega}_b}\qty[e^{-i(\omega_a-\omega_m)t} + e^{i(\omega_a + \omega_m)t}],$ $b_I(t) = b e^{-i\qty[\bar{\omega}_b t + \frac{\varepsilon}{\omega_m}\sin(\omega_m t)]}$. Here, the subindex $I$ denotes the change of frame. Treating the bath in the Markov approximation, we compute the evolution of the $\zeta$-mode density matrix as
\begin{equation}
\dot{\rho}^{\zeta}_I = -\tr_{\text{bath}}\qty[H^{\text{int}}_I(t),\int_0^{\infty}dt'\qty[H^{\text{int}}_I(t'),\rho_I(t)]].
\label{eq:double_commutator}
\end{equation} 
Factorizing $\rho_I(t)\simeq {\rho}^{\zeta}_I(t)\otimes{\rho}^{\text{bath}}_I,$ we expand the double commutator in \cref{eq:double_commutator} as a function of the $b$-mode correlation functions, which are computed neglecting backaction from the $\zeta$-mode. More precisely, we employ the quantum regression formula \cite{gardiner2004quantum} to obtain
\begin{equation}
\expval{b_I(t+\tau)b^{\dag}_I(t)} = \expval{b_I(t)b^{\dag}_I(t)}\exp[-\int_t^{t+\tau}dt' \qty(i \omega_b(t')+\frac{\kappa_b}{2})].
\label{eq:correlation_function}
\end{equation} 
Assuming that the external resonator is in a state very close to vacuum at all times, we approximate $\expval{b_I(t)b^{\dag}_I(t)}\simeq 1$, while the other three possible correlation functions being taken equal to zero. The validity of these various assumptions was verified with numerical simulations by plotting the corresponding expectation values obtained from the full time-dependent master equations. Next, we use the Jacobi-Anger expansion to expand \cref{eq:correlation_function} as
\begin{equation}
\expval{b_I(t)b^{\dag}_I(t')} \simeq e^{-\frac{\kappa_b}{2}(t-t')}e^{-i\bar{\omega}_b(t-t')}\sum_{n,n'=-\infty}^{\infty} J_n\qty(\frac{\varepsilon}{\omega_m})J_{n'}\qty(\frac{\varepsilon}{\omega_m}) e^{-i\omega_m (nt-n't')}.
\label{eq:correlation_function_2}
\end{equation} 
Inserting \cref{eq:correlation_function_2} in \cref{eq:double_commutator}, and retaining only the non-rotating terms, we obtain the following dissipative rates 
\begin{equation}
\begin{split}
\Gamma_{\downarrow} = \sum_{n=-\infty}^{\infty} \Bigg\{&\frac{\kappa_b\bar{g}^2 J^2_n\qty(\frac{\varepsilon}{\omega_m})}{(\omega_{\zeta}-\bar{\omega}_b + n\omega_m)^2 + \qty(\frac{\kappa_b}{2})^2}+\frac{\kappa_b\bar{g}^2\qty(\frac{\varepsilon}{4\bar{\omega}_b})^2 J^2_n\qty(\frac{\varepsilon}{\omega_m})}{(\omega_{\zeta}-\bar{\omega}_b + (n-1)\omega_m)^2 + \qty(\frac{\kappa_b}{2})^2}+\frac{\kappa_b\bar{g}^2\qty(\frac{\varepsilon}{4\bar{\omega}_b})^2 J^2_n\qty(\frac{\varepsilon}{\omega_m})}{(\omega_{\zeta}-\bar{\omega}_b + (n+1)\omega_m)^2 + \qty(\frac{\kappa_b}{2})^2}\\
&+\frac{2\kappa_b\bar{g}^2\qty(\frac{\varepsilon}{4\bar{\omega}_b}) J_n\qty(\frac{\varepsilon}{\omega_m})J_{n+1}\qty(\frac{\varepsilon}{\omega_m})}{(\omega_{\zeta}-\bar{\omega}_b - n\omega_m)^2 + \qty(\frac{\kappa_b}{2})^2}+\frac{2\kappa_b\bar{g}^2\qty(\frac{\varepsilon}{4\bar{\omega}_b}) J_n\qty(\frac{\varepsilon}{\omega_m})J_{n+1}\qty(\frac{\varepsilon}{\omega_m})}{(\omega_{\zeta}-\bar{\omega}_b - (n+1)\omega_m)^2 + \qty(\frac{\kappa_b}{2})^2}+\frac{2\kappa_b\bar{g}^2\qty(\frac{\varepsilon}{4\bar{\omega}_b})^2 J_n\qty(\frac{\varepsilon}{\omega_m})J_{n+2}\qty(\frac{\varepsilon}{\omega_m})}{(\omega_{\zeta}-\bar{\omega}_b - (n+1)\omega_m)^2 + \qty(\frac{\kappa_b}{2})^2}\Bigg\},
\end{split}
\label{eq:gamma_down_full}
\end{equation} 
and 
\begin{equation}
\begin{split}
\Gamma_{\uparrow} = \sum_{n=-\infty}^{\infty} \Bigg\{&\frac{\kappa_b\bar{g}^2 J^2_n\qty(\frac{\varepsilon}{\omega_m})}{(\omega_{\zeta}+\bar{\omega}_b - n\omega_m)^2 + \qty(\frac{\kappa_b}{2})^2}+\frac{\kappa_b\bar{g}^2\qty(\frac{\varepsilon}{4\bar{\omega}_b})^2 J^2_n\qty(\frac{\varepsilon}{\omega_m})}{(\omega_{\zeta}+\bar{\omega}_b - (n+1)\omega_m)^2 + \qty(\frac{\kappa_b}{2})^2}+\frac{\kappa_b\bar{g}^2\qty(\frac{\varepsilon}{4\bar{\omega}_b})^2 J^2_n\qty(\frac{\varepsilon}{\omega_m})}{(\omega_{\zeta}+\bar{\omega}_b - (n-1)\omega_m)^2 + \qty(\frac{\kappa_b}{2})^2}\\
&+\frac{2\kappa_b\bar{g}^2\qty(\frac{\varepsilon}{4\bar{\omega}_b}) J_n\qty(\frac{\varepsilon}{\omega_m})J_{n+1}\qty(\frac{\varepsilon}{\omega_m})}{(\omega_{\zeta}+\bar{\omega}_b + n\omega_m)^2 + \qty(\frac{\kappa_b}{2})^2}+\frac{2\kappa_b\bar{g}^2\qty(\frac{\varepsilon}{4\bar{\omega}_b}) J_n\qty(\frac{\varepsilon}{\omega_m})J_{n+1}\qty(\frac{\varepsilon}{\omega_m})}{(\omega_{\zeta}+\bar{\omega}_b + (n+1)\omega_m)^2 + \qty(\frac{\kappa_b}{2})^2}+\frac{2\kappa_b\bar{g}^2\qty(\frac{\varepsilon}{4\bar{\omega}_b})^2 J_n\qty(\frac{\varepsilon}{\omega_m})J_{n+2}\qty(\frac{\varepsilon}{\omega_m})}{(\omega_{\zeta}+\bar{\omega}_b + (n+1)\omega_m)^2 + \qty(\frac{\kappa_b}{2})^2}\Bigg\}.
\end{split}
\label{eq:gamma_up_full}
\end{equation}
Maximizing each of the summands in \cref{eq:gamma_down_full} and \cref{eq:gamma_up_full}, and discarding all but most significant terms, these expressions reduce to the forms $\Gamma_{\downarrow} = \frac{4g'^{2}}{\kappa_b}$ and $\Gamma_{\uparrow} = \frac{4g'^{2}}{\kappa_b}\Big/{\qty[\qty(\frac{2\omega_{\zeta}}{\kappa_b/2})^2+1]}$ quoted in the main text of the article. We note that the interaction of the $\zeta$-mode with the $b$-mode also leads to a frequency shift which can be taken into account by a slight renormalization of the modulation frequency $\omega_m$. This effect is negligible for the parameter regime considered here.

\section*{References}
\bibliographystyle{iopart-num}
\bibliography{library}

\end{document}